\documentclass[pre,aps,reprint,amsmath,amssymb,superscriptaddress]{revtex4-1}
\usepackage[utf8]{inputenc}
\usepackage{graphicx}
\usepackage{amsmath,amssymb}
\usepackage[dvipsnames]{xcolor}
\usepackage[colorlinks=true,citecolor=ForestGreen,linkcolor=Red,urlcolor=Blue,hypertexnames=false]{hyperref}
\usepackage[sort&compress]{natbib}
\newcommand{\ie}{\emph{i.e.}}
\newcommand{\eg}{\emph{e.g.}}
\newcommand{\avg}[1]{\langle #1\rangle}
\newcommand{\eref}[1]{Eq.~(\ref{#1})}

\newcommand{\abs}[1]{\lvert #1 \rvert}
\newcommand{\babs}[1]{\big\lvert #1 \big\rvert}
\newcommand{\vek}[1]{\boldsymbol{#1}}
\newcommand{\figSI}[2]{\includegraphics[scale=#1]{#2}}
\begin{document}
\title{The fragility of opinion formation in a complex world}
\author{Matúš Medo}
\email{matus.medo@unifr.ch}
\affiliation{Institute of Fundamental and Frontier Sciences, University of Electronic Science and Technology of China, Chengdu 610054, PR China}
\affiliation{Department of Radiation Oncology, Inselspital, University Hospital of Bern, and University of Bern, 3010 Bern, Switzerland}
\affiliation{Department of Physics, University of Fribourg, 1700 Fribourg, Switzerland}
\author{Manuel S. Mariani}
\affiliation{Institute of Fundamental and Frontier Sciences, University of Electronic Science and Technology of China, Chengdu 610054, PR China}
\affiliation{URPP Social Networks, Universit\"at Z\"urich, Switzerland}
\author{Linyuan L\"u}
\affiliation{Institute of Fundamental and Frontier Sciences, University of Electronic Science and Technology of China, Chengdu 610054, PR China}
\affiliation{Alibaba Research Center for Complexity Sciences, Hangzhou Normal University, Hangzhou 311121, PR China}
\affiliation{Beijing Computational Science Research Center, Beijing 100193, PR China}
\date{\today}

\begin{abstract}
With vast amounts of high-quality information at our fingertips, how is it possible that many people believe that the Earth is flat and vaccination harmful? Motivated by this question, we quantify the implications of an opinion formation mechanism whereby an uninformed observer gradually forms opinions about a world composed of subjects interrelated by a signed network of mutual trust and distrust. We show numerically and analytically that the observer's resulting opinions are highly inconsistent (they tend to be independent of the observer's initial opinions) and unstable (they exhibit wide stochastic variations). Opinion inconsistency and instability increase with the world complexity represented by the number of subjects, which can be prevented by suitably expanding the observer's initial amount of information. Our findings imply that even an individual who initially trusts credible information sources may end up trusting the deceptive ones if at least a small number of trust relations exist between the credible and deceptive sources.
\end{abstract}

\maketitle

\section*{Introduction}
Identifying potential mechanisms behind the formation of opinions in society is vital to understand how polarization emerges in society~\cite{baumann2020modeling}, how misinformation spreads and can be prevented~\cite{del2016spreading}, and how science can be effectively communicated to the public~\cite{scheufele2019science}. Despite recent advances in how opinions propagate in social networks~\cite{baumann2020modeling}, how artificial agents promote low-credibility content in social media~\cite{shao2018spread,stella2018bots} and how rapidly misinformation spreads compared to reliable content~\cite{del2016spreading,vosoughi2018spread}, misinformation still thrives in our society. This is well-exemplified by the recent growth of anti-vaccination views and the related existence of antivaccination clusters in online social networks~\cite{johnson2020online}. The popularity of unreliable opinions -- which is especially dangerous during global emergencies, such as the recent COVID-19 pandemic~\cite{yang2020prevalence} -- calls for a deeper investigation of the possible drivers behind the process whereby individuals form opinions in a society.

We show that, even in the absence of a social network of influence, inconsistent and unstable opinions can emerge from a process whereby an uninformed individual forms opinions about a world composed of interrelated subjects. Existing models of opinion formation \cite{degroot1974reaching,friedkin1990social,castellano2009statistical,sirbu2017opinion,baumann2020modeling} and cultural dynamics~\cite{axelrod1997dissemination} focus on opinion or culture propagation on a social network of influence. Departing from existing models, we develop a modeling approach that focuses on the process whereby an individual observer forms opinions about a set of interrelated subjects. In existing models~\cite{friedkin1990social,castellano2009statistical,sirbu2017opinion}, an individual can form opinions on distinct topics as the result of independent realizations of an opinion propagation process on a social network of influence. By contrast, the proposed model takes into account the connections among topics~\cite{lobato2014examining}, which we find to be a key determinant of opinion inconsistency.

Inspired by Heider's social balance theory~\cite{heider1946attitudes,cartwright1956structural,wasserman1994social}, and its validation on data on armed conflicts among countries~\cite{moore1979structural,crescenzi2007reputation} and large-scale social media~\cite{facchetti2011computing,zheng2015social,lerner2020free}, our model assumes that an individual observer gradually forms opinions on a set of subjects connected by signed links representing positive and negative relations, respectively. The subjects on which the opinions are formed can represent governments, politicians, news media, or other individuals that belong to two different camps. While such systems tend to form macroscopic structures such as two opposing camps~\cite{moore1979structural,adamic2005political,marvel2011continuous}, these structures are generally imperfect. Two countries, for example, can belong to the same alliance whose members generally have positive relations, yet their mutual relation can be negative due to historic or economic reasons (consider the two NATO members, Greece and Turkey, and their long-term issues). In science, it has repeatedly occurred that a Nobel prize recipient endorsed conspiracy theories, as was recently the case with Luc Montagnier's controversial claims on the origin of the ongoing COVID-19 pandemic.

Forming a reliable opinion about a complex subject requires effortful reasoning. However, psychological research indicates that humans tend to be rather driven by simple heuristics when forming opinions about complex topics, sometimes reaching opinions that violate basic logic rules~\cite{tversky1983extensional,kahneman2011thinking}. The limitations of our cognition have important consequences. For example, the susceptibility to partisan fake news was recently found to be driven more by ``lazy reasoning'' than by partisan bias~\cite{pennycook2019lazy}. This motivates us to study the problem of an observer who starts with opinions on a small set of subjects (seed opinions) and applies a local rule (heuristics) to form opinions on the remaining subjects by relying only on their explicit signed relations.

We find that even a small fraction of misleading links in the relation network (\eg, a link of mutual trust between a scientific and low-credibility information source) leads to the resulting opinions that are both inconsistent with the observer's seed opinions and vary significantly between model realizations. We determine analytically the relation between average opinion consistency and the world complexity, represented by the number of subjects, which demonstrates that opinion consistency grows as the world complexity increases. This increase can be prevented by suitably increasing the observer's initial number of independent opinions. Although opinion consistency depends on network topology and can be improved by considering a more sophisticated local opinion formation mechanism, our main conclusions are robust to variations of the network topology and the opinion-formation mechanism.

Our findings point to the inherent fragility of the opinion formation process in a world composed of many interrelated subjects and, at the same time, suggest strategies to increase its reliability. Since subjects may represent co-existing scientific or low-credibility information sources, our model presents a contributing mechanism for how misinformation sources may gain their audience.

\section*{Results}

\subsection{Opinion formation model.}
\begin{figure*}
\includegraphics[scale=0.68]{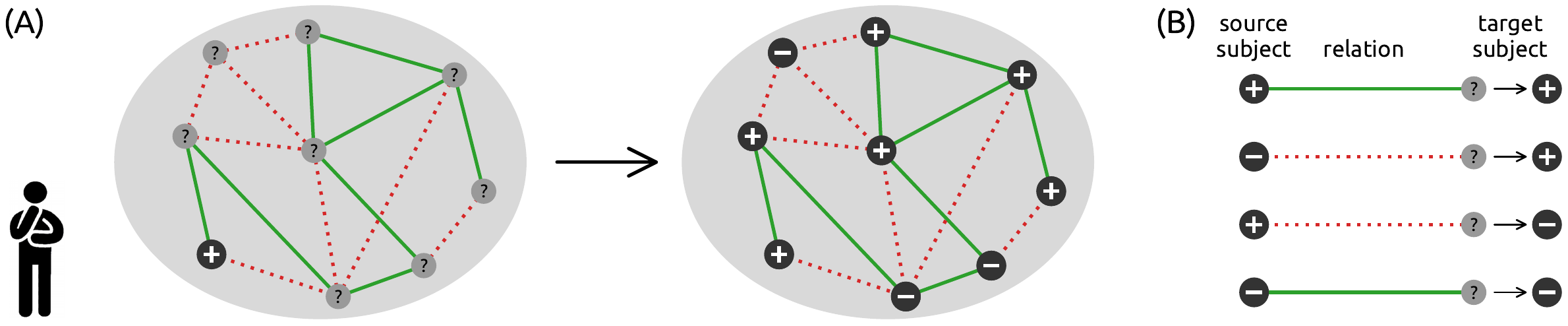}
\caption{\label{fig:model_scheme} The opinion formation model. (A) Starting from a small set of seed opinions and a world of unknown subjects (gray nodes), an observer gradually forms opinions (black circles, $\pm$) on all the subjects. The subjects are interconnected by mutual relations of trust (solid green lines) or distrust (dotted red lines). (B) The formed opinion is determined as a product of the opinion on the source subject and the sign of the relation between the source subject and the target subject. A positive opinion is formed when the source opinion and the relation are both positive or both negative; a negative opinion is formed otherwise.}
\end{figure*}

We consider an individual observer who gradually develops opinions on a world composed of $N$ interrelated subjects (see Fig.~\ref{fig:model_scheme}A). The number of subjects represents the complexity of the world. Each opinion is for simplicity assumed to take one of three possible states: no opinion, a positive opinion (trust), or a negative opinion (distrust). The observer's opinions can be formally represented by an $N$-dimensional opinion vector $\vek{o}$ whose element $o_i$ represents the opinion on subject $i$; $o_i\in\{-1,0,1\}$ corresponds to a negative opinion, no opinion, and a positive opinion, respectively. The subjects form a signed undirected network of relations. These relations are represented by a symmetric $N\times N$ relation matrix whose element $R_{ij}$ represents the trust relation between subjects $i$ and $j$; $R_{ij}\in\{-1,0,1\}$ corresponds to a negative relation, no relation, and a positive relation, respectively. We emphasize the main difference between this setting and traditional opinion formation models based on propagation on networks of social influence~\cite{degroot1974reaching,friedkin1990social,castellano2009statistical,sirbu2017opinion}: in existing models, simulating the opinion formation on $N$ subjects would require running $N$ \emph{independent} realizations of the opinion formation process, which would miss the \emph{interconnectedness} among subjects; by contrast, in the proposed approach, the interconnectedness among subjects is naturally encoded in the relation matrix $\mathsf{R}$.

The observer's opinion formation starts from an initial condition where the observer has an initial opinion on a ``seed'' subset of subjects, $\mathcal{S}$ (seed opinions). The observer then gradually forms an opinion on each of the remaining subjects via sequential opinion formation, until opinions on all subjects are formed. Once formed, the opinions are not updated. In one step, a target subject $i$ is chosen at random from the pool of subjects with no opinion ($o_i=0$). The observer then attempts to form an opinion on $i$. From all subjects $j$ with an opinion ($o_j\neq 0$) that are adjacent to $i$ ($R_{ij}\neq 0$), we choose one subject at random (source subject). The opinion $o_i$ is then set to $o_j R_{ji}$ (see Fig.~\ref{fig:model_scheme}B). As a result, a positive opinion on $i$ is formed if either: (1) the observer has a positive opinion on $j$ and the relation between $j$ and $i$ is positive (``the friends of my friends are also my friends'') or (2) the observer has a negative opinion on $j$ and the relation between $j$ and $i$ is negative (formalizing the ancient proverb ``the enemies of my enemies are my friends''). A negative opinion on $i$ is formed otherwise. Note that this mechanism produces a balanced triad consisting of the observer and subjects $i$ and $j$ (in Heider's original sense of heterogeneous triads that can include both individuals as well as objects~\cite{heider1946attitudes,van2011micro}). The observer then continues with a next subject until opinions on all subjects have been formed. This opinion formation process---which we refer to as the random neighbor rule as it forms opinions using neighboring subjects chosen at random---is purposely simple as it intends to imitate an observer with limited cognitive resources (see~\cite{pennycook2019lazy} for a recent account on susceptibility to fake news driven by ``lack of reasoning''). We study a more thorough process (majority rule) below.

The opinion formation outcome is not deterministic (except for the special case when all paths in the subject network are balanced; see section S1 in Supporting Material, SM) as it is influenced by the order in which subjects are chosen for opinion formation as well as the source subject choices. For a given relation network and a set of seed opinions, individual realizations of the process correspond to a population of independent individuals or, alternatively, various possible ``fates'' of a single individual. We study outcomes of multiple model realizations to characterize statistical properties of the resulting opinions. For simulations on synthetic relation networks, we additionally average over various network realizations to remove possible effects of a specific network topology on the results.

\subsection{Opinion formation simulations on synthetic networks.}
We now study the opinion formation model on a specific relation network where the subjects form two camps. This scenario is relevant to various real situations~\cite{moore1979structural,adamic2005political,marvel2011continuous}: The two camps can represent two opposing political parties (such as democrats and republicans), standard news outlets and false news outlets, or scientists and conspiracy theorists, for example. In synthetic networks, each camp consists of $N/2$ subjects. Every subject is connected by signed links with $z$ random subjects, thus creating a random network of trust with average degree $z$. If subjects from the same camp are linked, the sign of their relation is $+1$ with probability $1-\beta$ and $-1$ otherwise. Similarly, if subjects from different camps are linked, the sign of their relation is $-1$ with probability $1-\beta$ and $+1$ otherwise. Parameter $\beta\in[0, 0.5]$ thus plays the role of structural noise. As $\beta$ grows, the negative relations become more common within each camp and positive relations become more common across the camps. When $\beta=0.5$, the two camps become indistinguishable by definition. The network's level of structural balance~\cite{cartwright1956structural,harary1959measurement} is the ratio of the number of balanced triads to all triads in the network. In our case,
\begin{equation}
\label{B}
B=(1-\beta)^3 + 3(1-\beta)\beta^2,
\end{equation}
which corresponds to either all links of a triad (the first term) or one link of a triad (the second term) respecting the two-camp structure, producing a balanced triad as a result. $B$ grows monotonously with $\beta$. The equation can be inverted, yielding $\beta=(1+\sqrt[3]{1-2B})/2$, which can be used to write our results in terms of $B$ instead of $\beta$.

We assume that the observer has initially a positive opinion on $N_S$ seed subjects from camp 1, and we examine whether the observer ends up with a positive opinion on other subjects from camp 1 and a negative opinion on subjects from camp 2, or not. If the two camps represent scientists and conspiracy theorists, for example, the corresponding practical question is whether an observer who initially trusts a scientist would end up predominantly trusting scientists or conspiracy theorists. Without noise ($\beta=0$), the opinion formation leads to a definite outcome: A positive opinion on all subjects from camp 1 and a negative opinion on all subjects from camp 2. In such a case, we say that the opinions are perfectly consistent with the underlying two-camp structure of the relationship network among the subjects. Opinion consistency of a resulting opinion vector, $\boldsymbol{o}$, can be measured as
\begin{equation}
\label{consistency}
C(\boldsymbol{o},\boldsymbol{T}) = \frac1{N-N_S}\sum_{j\not\in\mathcal{S}} o_j T_j
\end{equation}
where $\mathcal{S}$ is the set of seed subjects and $\boldsymbol{T}$ represents the ground-truth structure of the relation network (in our case, $T_j=1$ for $j$ from camp 1 and $T_j=-1$ for $j$ from camp 2). If the observer's opinions are chosen at random, the resulting consistency is zero on average. A zero or small consistency value thus indicates that the observer's opinions are independent of the seed opinion and thus inconsistent with the two-camp structure of the relationship network. Negative consistency is also possible: The observer starts with a positive opinion on subjects from camp 1 but ends with more positive opinions in camp 2 than in camp 1.

\begin{figure*}
\centering
\includegraphics[scale=0.68]{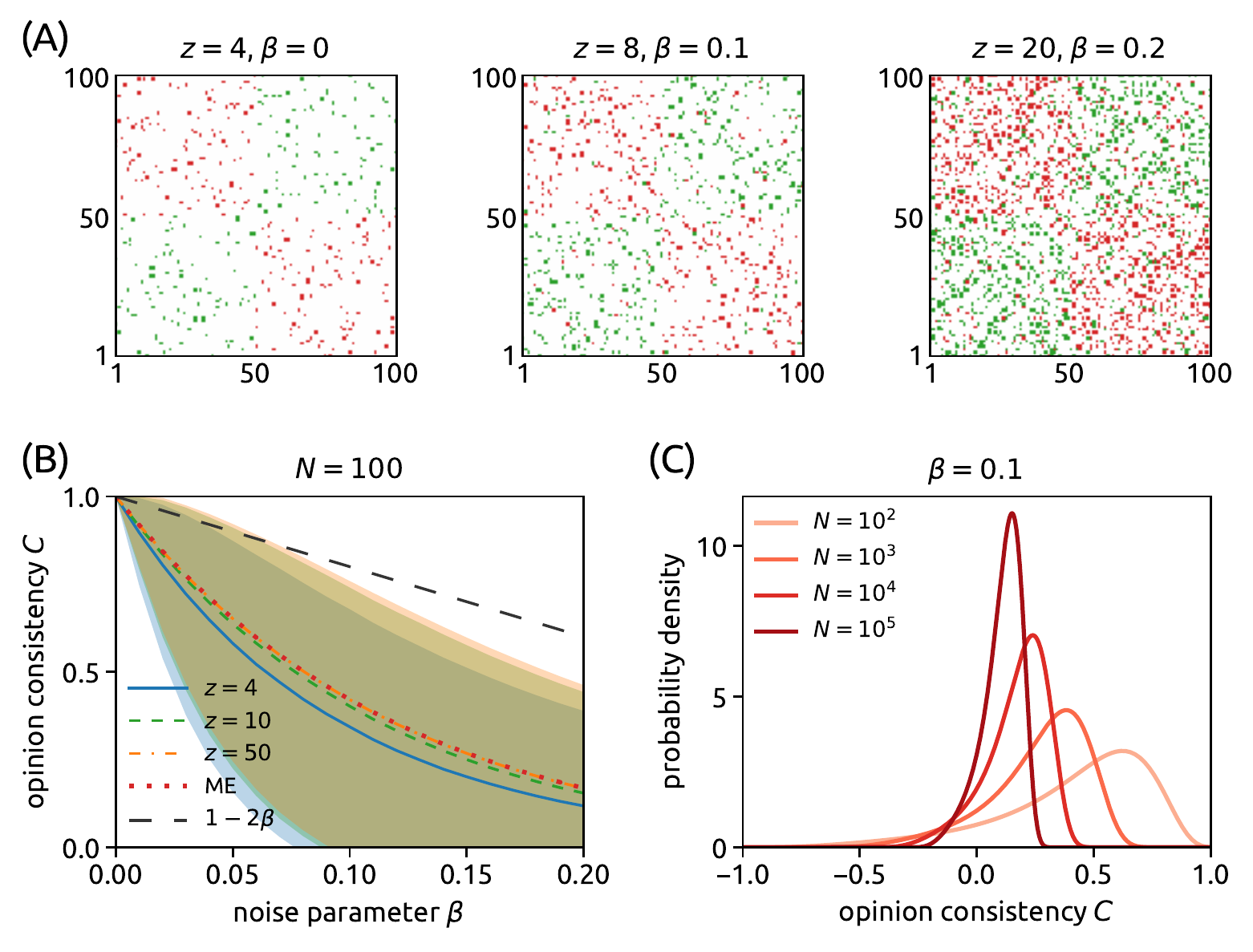}
\caption{\label{fig:results1}
Opinion formation on random relation networks with a two-camp structure.
(A) Examples of random relation networks with a two-camp structure for $100$ subjects and various values of mean degree, $z$, and structural noise, $\beta$. Subjects 1--50 are from camp 1 and subjects 51--100 are from camp 2. The green and red points represent positive and negative relations, respectively.
(B) Opinion consistency for $N=100$ subjects, one seed opinion, and various values of $z$. The lines show mean values and the shaded regions show the 10th--90th percentile ranges (both computed from 1000 model realizations for each of 1000 realizations of the relation network). The dotted line labeled ME shows the mean consistency computed by numerically solving the master equation given by \eref{ME}. The dark dashed line shows the expected consistency if its relationship with the level of noise was linear [$C_0(\beta)=1-2\beta$].
These results demonstrate that consistency decreases quickly with structural noise.
(C) The consistency distributions obtained using \eref{ME} for $\beta=0.1$, $N_S=1$ and a growing number of subjects, $N$. As $N$ increases, the distributions shrink and their peaks shift toward zero.}
\end{figure*}

Knowing that opinion consistency is one in the absence of noise, how does it change as the noise parameter $\beta$ grows? Numerical simulations for a set of $100$ subjects and one seed opinion show (Fig.~\ref{fig:results1}A) that opinion consistency decreases rapidly with $\beta$. Indeed, if the relationship between consistency and noise was linear, we would have expected $C_0(\beta):=1-2\beta$ which starts at one when $\beta=0$ and reaches zero when $\beta=0.5$ as the two camps then cannot be distinguished by definition. By contrast, we observe a substantially faster decay of the mean consistency $\mu_C(\beta)$.
In addition, the consistency values vary strongly between model realizations. For $\beta=0.02$, for example, mean consistency is only $0.80$ and there are model realizations with consistency below $0.54$ and above $0.97$ (the 10th and 90th percentile, respectively, of the obtained consistency values for $z=4$). This means that even when the noise is small, some sets of formed opinions are in a dramatic disagreement with the observer's seed opinion. To appreciate the level of noise in real data, Moore~\cite{moore1979structural} reported that 80\% of triads among middle East countries are balanced. \eref{B} shows that such a level of structural balance is achieved at $\beta\approx0.08$ in our two-camp networks. In Fig.~\ref{fig:results1}B, mean opinion consistency at $\beta=0.08$ is as low as $0.42$ (for $z=10$). These results confirm our initial hypothesis that a realistic level of noise leads to the adoption of a large fraction of opinions that do not align with the observer's initial opinion.

\subsection{Master equation for opinion consistency and its solution.}
The opinion formation with the two-camp structure can be studied analytically under the assumption of homogeneous mixing~\cite{pastor2015epidemic}. It is advantageous to study the problem in terms of the number of formed opinions, $n$, and the number of \emph{consistent opinions}, $c$ (that is, the opinions that are consistent with the seed opinions and the two-camp structure). By rewriting the sum $\sum_{j\neq i} o_jT_j$ in \eref{consistency} as $2c-N+N_S$, we obtain opinion consistency as $C=(2c-N+N_S)/(N-N_S)$.

When the observer forms a new opinion, $n$ increases by one and $c$ either increases by one (if the new opinion is consistent) or remains constant. We introduce the probability distribution of $c$ when $n$ opinions have been formed, $P(c; n)$, for which the master equation (see Methods for the derivation) has the form
\begin{widetext}
\begin{equation}
\label{ME}
P(c; n) = P(c - 1; n - 1)\,\frac{c(1-2\beta) + \beta(n+1)-1}{n - 1} + P(c; n - 1) \bigg[1 - \beta - \frac{c(1-2\beta)}{n-1}\bigg]
\end{equation}
\end{widetext}
The initial condition $P(N_S;N_S)=1$ represents that all $N_S$ seed opinions are consistent. \eref{ME} can be solved numerically and the obtained solution $P(c;n)$ can be used to compute the corresponding mean opinion consistency. The numerical solution agrees well with the model simulations (Fig.~\ref{fig:results1}B), in particular when the relation network is not sparse ($z\gtrsim 10$).

\begin{figure*}
\centering
\includegraphics[scale=0.68]{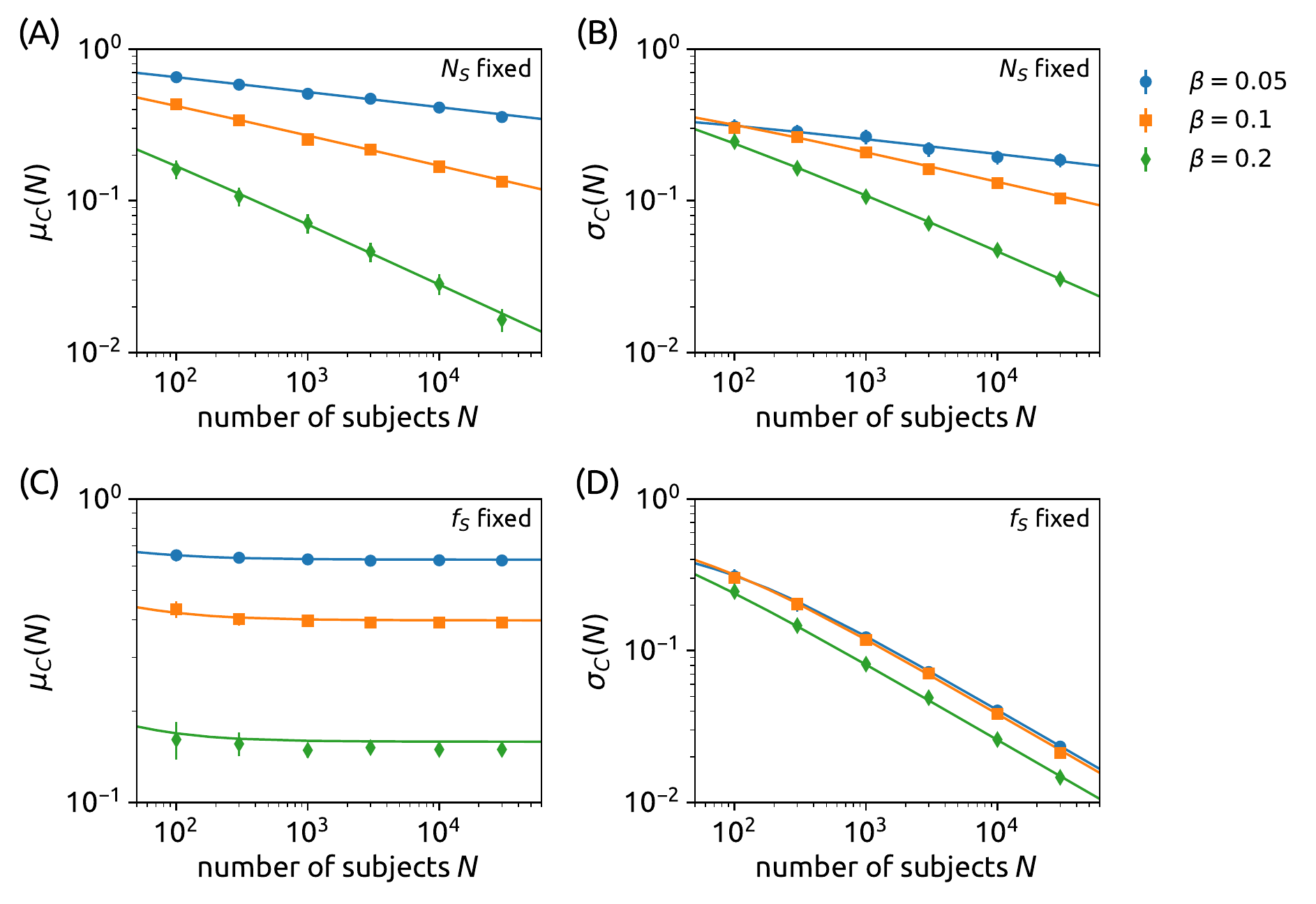}
\caption{Opinion consistency for synthetic worlds. Mean opinion consistency, $\mu_C$, and the standard deviation of consistency, $\sigma_C$, as functions of the number of subjects, $N$, for $z=50$. The symbols show the results obtained by running the opinion formation model on 1000 relation network realizations; the error bars are too small to be shown. The solid lines show $\mu_C(N)$ and $\sigma_C(N)$ obtained by solving the master equation (\eref{mu_n} and Eq.~(S6) in Supporting Material). (A,B) Results for a single seed opinion. In this scenario, both $\mu_C$ and $\sigma_C$ converge to zero as $N$ grows: there is a tension between opinion consistency and the world's complexity. (C,D) Results for the number of seed opinions, $N_S$, given by $N_S=f_SN$ (here $f_S=0.01$). In this scenario, the tension between consistency and complexity is prevented: $\mu_C$ converges to $f_S^{2\beta}$ (panel C) and $\sigma_C$ converges to zero (panel D).}
\label{fig:scaling}
\end{figure*}

\eref{ME} allows us to investigate the dependence between opinion consistency and the world complexity, represented by the number of subjects, $N$. A surprising finding is that as the number of subjects increases, the distribution of $C$ obtained by solving \eref{ME}, $P(C)$, does not approach a well-defined limit distribution, but instead steadily shifts towards $C=0$ and becomes narrower in the process (Figure~\ref{fig:results1}C). We study this behavior by computing the mean opinion consistency, $\mu_C(N)$, and the standard deviation of consistency, $\sigma_C(N)$. 

Multiplying \eref{ME} with $c$ and summing it over $c=N_S,\dots,N$ yields the recurrence equation
\begin{equation}
\label{recurr}
\avg{c(n)} = \frac{n - 2\beta}{n-1}\avg{c(n-1)} + \beta
\end{equation}
with the initial condition $\avg{c(N_S)} = N_S$ (the seed opinions are assumed to be correct). This recurrence equation can be solved in general, yielding
\begin{equation}
\label{Cn}
\avg{c(N)} = \frac12\bigg[N+\frac{\Gamma(N+1-2\beta)\Gamma(N_S+1)}{\Gamma(N_S+1-2\beta)\Gamma(N)}\bigg].
\end{equation}
For $N_S=1$, the corresponding mean consistency is
\begin{equation}
\label{mu_n}
\mu_C(N) = \bigg[\frac{\Gamma(N+1-2\beta)}{2\Gamma(2-2\beta)\Gamma(N)}-1\bigg] / (N - 1)
\end{equation}
with the leading contribution
\begin{equation}
\mu_{C}(N) = N^{-2\beta}/\Gamma(2-2\beta) + O(1/N).
\end{equation}
This shows that the mean opinion consistency vanishes in the limit $N\to\infty$. The leading-term contribution to $\sigma_C(N)$ is also proportional to $N^{-2\beta}$ when $\beta\leq 1/4$. When $\beta>1/4$, the leading term becomes proportional to $N^{-1/2}$. These analytic results agree with numerical simulations of the model (Figure~\ref{fig:scaling}).

The behavior demonstrated by Figures~\ref{fig:results1}B and \ref{fig:scaling}, and supported by the analytic solution above, has important consequences. It shows that as the world complexity increases, the formed opinions become on average less consistent with the seed opinions and the two-camp structure of the subject network. Crucially, the opinion consistency is zero in the limit of an infinite number of subjects for any positive level of noise, $\beta$, in the subject relation network: in the limit of an infinite-complexity world, even a tiny amount of noise is enough to nullify opinion consistency. These results are robust with respect to variations of the relation network structure and using more than one seed opinion (see Sec.~S3 in SM).

The convergence of opinion consistency to zero as $N\to\infty$ can be avoided if the number of seed opinions grows linearly with $N$ so that the fraction of seed opinions remains constant. Assuming that $N_S=f_SN$, \eref{Cn} can be used to show that the mean consistency approaches to
\begin{equation}
\mu_C=f_S^{2\beta}
\end{equation}
in the limit $N\to\infty$ and the standard deviation of consistency vanishes as $1/\sqrt{N}$ (see Sec.~S2 SM). This scaling relation determines the necessary proportion of seed opinions, $f_S$, needed to achieve a desired opinion consistency, $\mu_C$, for given $\beta$. These results are confirmed by numerical simulations shown in Fig.~\ref{fig:scaling}C,D. Despite having a positive limit value, opinion consistency still decreases quickly with noise in the relationship network when $f_S$ is small (see Fig.~S2 in SM).

\begin{figure*}
\includegraphics[scale=0.68]{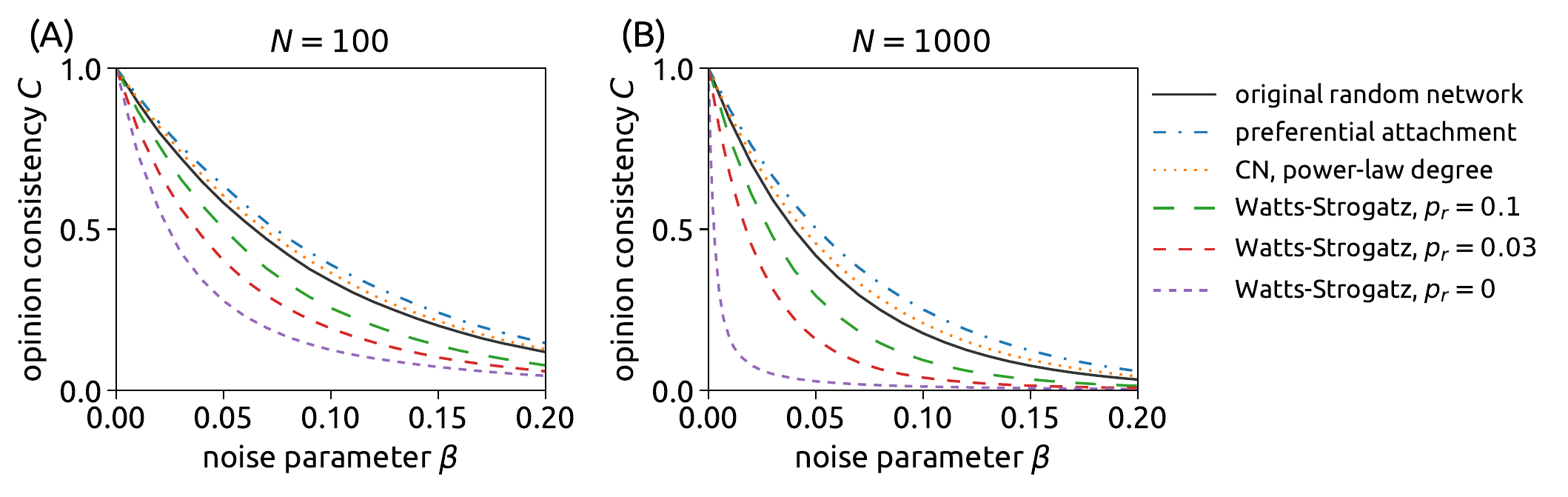}
\caption{Mean opinion consistency for various topologies of synthetic relation networks: the dependence on $\beta$ for (A) $N = 100$ and (B) $N = 1,000$. The results are averaged over 1,000 model realizations on each of 1,000 network realizations; $z\approx 4$ for every network topology.}
\label{fig:topology_beta}
\end{figure*}

While our findings hold qualitatively when a different topology of the relation network is used, the mean opinion consistency values are heavily affected by the network topology (see Figure~\ref{fig:topology_beta}). We run the opinion formation model on a growing preferential attachment network, a configuration model (CN) network with a power-law degree distribution, and Watts-Strogatz networks with various values of the rewiring probability, $p_r$ (see Sec.~S3.3 in SM for details on the network construction). We find that networks with broad degree distributions lead to higher opinion consistency which decays with $N$ slower (see Fig.~S5 in SM) than in the previously studied random networks. By contrast, Watts-Strogatz networks yield lower opinion consistency which further decreases as the networks become more regular through lowering the rewiring probability, $p_r$.

\subsection{Opinion formation using the majority rule.}
The results described above hold for the opinion formation model where a random neighbor of a target subject is chosen as the reference. We chose this model to study the consequences of a cognitively easy opinion formation model. At this stage, one might object that the observed sensitivity of opinion consistency to noise might be because each formed opinion directly relies on only one previously formed opinion, and it might disappear if the observer incorporates the information from more neighbors before forming an opinion. To rule out this potential argument, we investigate a model where \emph{all} neighbors of a target subject are considered before forming the opinion. Denote the numbers of neighbors leading to the adoption of a positive and a negative opinion (determined as in Figure~\ref{fig:model_scheme}B) as $n_P$ and $n_N$, respectively. If $n_P>n_N$, the observer forms a positive opinion. If $n_N>n_P$, the observer forms a negative opinion. If $n_P=n_N$, a random opinion is formed. We refer to this as the majority opinion formation rule. It is more demanding than the original random neighbor rule based on choosing a random neighbor as it assumes that the observer carefully collects \emph{all} evidence for forming an opinion on a target subject. The majority rule is nevertheless still a local rule as it only considers direct neighbors of a target node.

\begin{figure*}
\includegraphics[scale=0.68]{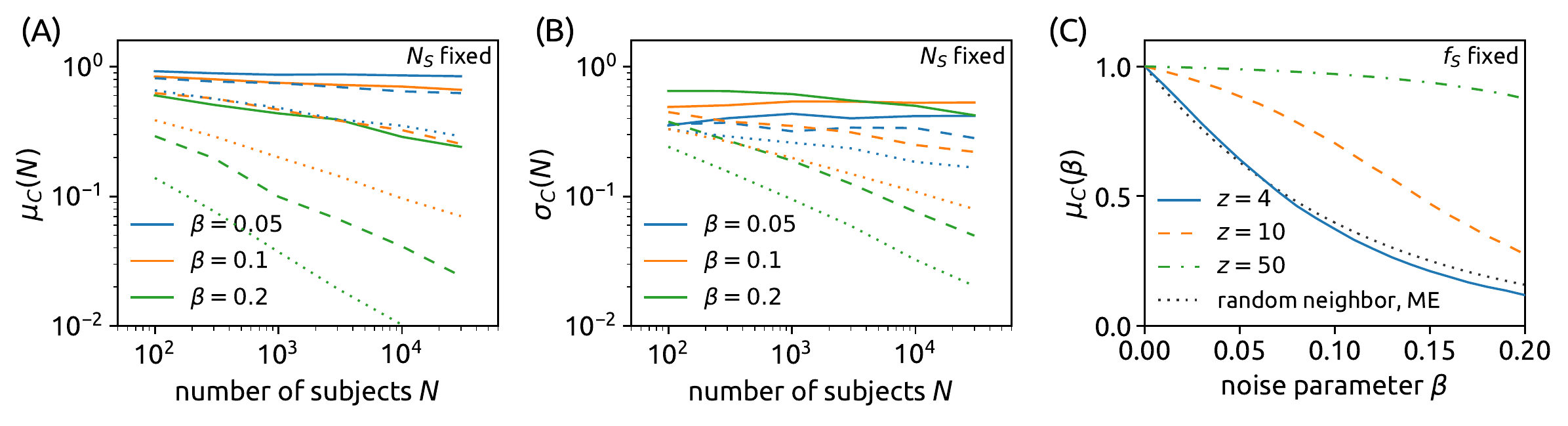}
\caption{Opinion consistency achieved in synthetic worlds by the majority rule. (A, B) The dependencies of $\mu_C(N)$ and $\sigma_C(N)$ on $N$ are affected by both $\beta$ and $z$ (the solid, dashed, and dotted lines show results for $z$ equal 50, 10, and 4, respectively). Here $N_S=1$ is fixed.
(C) When $N_S=f_S N$, $\mu_C(N)$ converges to a positive value and $\sigma_C(N)$ vanishes as $N$ grows. We plot here $\mu_C$ as a function of the structural noise, $\beta$, for $f_S=0.01$ and $N=10,000$ (results are averaged over model realizations on 1,000 independent synthetic networks). The dotted line shows the analytical result $\mu_C=f_S^{2\beta}$ for the random neighbor rule when a fixed fraction of seed opinions, $f_S$, is used.}
\label{fig:majority}
\end{figure*}

Using the majority rule, a scaling analogous to Figure~\ref{fig:scaling} can be observed (Figure~\ref{fig:majority}A,B). The important difference is that the scaling exponent now depends on both $\beta$ and $z$ whereas a higher mean degree, $z$, generally leads to $\mu_C(N)$ and $\sigma_C(N)$ decaying slower with $N$. Except for the smallest used noise and the highest used degree ($\beta=0.05$ and $z=50$), all fitted slopes are significantly positive. Since the majority rule does not lend itself to analytical computation, whether the limit of $\mu_C(N)$ is indeed positive when $\beta$ is sufficiently small and $z$ is sufficiently high remains an open question. Figure~\ref{fig:majority}C shows results for a fixed fraction of seed nodes. We see that when the network density is low ($z=4$), the majority rule achieves results that are comparable to those of the random neighbor rule. When $z$ increases, the majority rule leads to significantly more consistent opinions than the random neighbor rule. It has to be noted, though, that when $z$ is large, the cost for the observer to collect and analyze all information for opinion-making is large too.

\subsection{Opinion formation simulations on real networks.}
The trust consistency metric requires information on the ground truth structure of the relation network (such as the assignment of subjects to one of the two camps in the case of a two-camp structure). Before analyzing empirical data, we aim to introduce a proxy for opinion consistency that does not require such information which is typically not available for real data. To this end, we introduce \emph{opinion stability}, $S$, which measures the extent to which elements of the opinion vector are the same in independent realizations of the opinion formation model (see Methods for the definition). If an opinion on a given subject always ends up positive (or always negative), it is a sign of a robust opinion and it contributes positively to opinion stability. Small opinion stability indicates that the opinion formation outcomes are highly volatile and, in turn, they do not comply with the division of subjects in camps in synthetic networks. It can be shown that when the relation network's level of structural balance is one, opinion stability is one as well.

In synthetic worlds, the opinion stability metric behaves as required when the relation network is sufficiently dense ($z\gtrsim 10$): $S=1$ in synthetic networks when $\beta=0$ and $S$ is close to zero when $\beta=0.5$ (see Sec.~S4 in SM). In fact, the values of opinion consistency and opinion stability are nearly the same for all $\beta$ values. The main reason for this agreement between stability, $S$, and consistency, $C$, is that high opinion consistency can be only achieved when the opinions in question are the same in all model realizations which in turn leads to high opinion stability. Crucially, opinion stability vanishes as the number of subjects grows to infinity similarly as we have seen it for opinion consistency (see Fig.~S8 in SM).

\begin{figure*}
\centering
\includegraphics[scale=0.68]{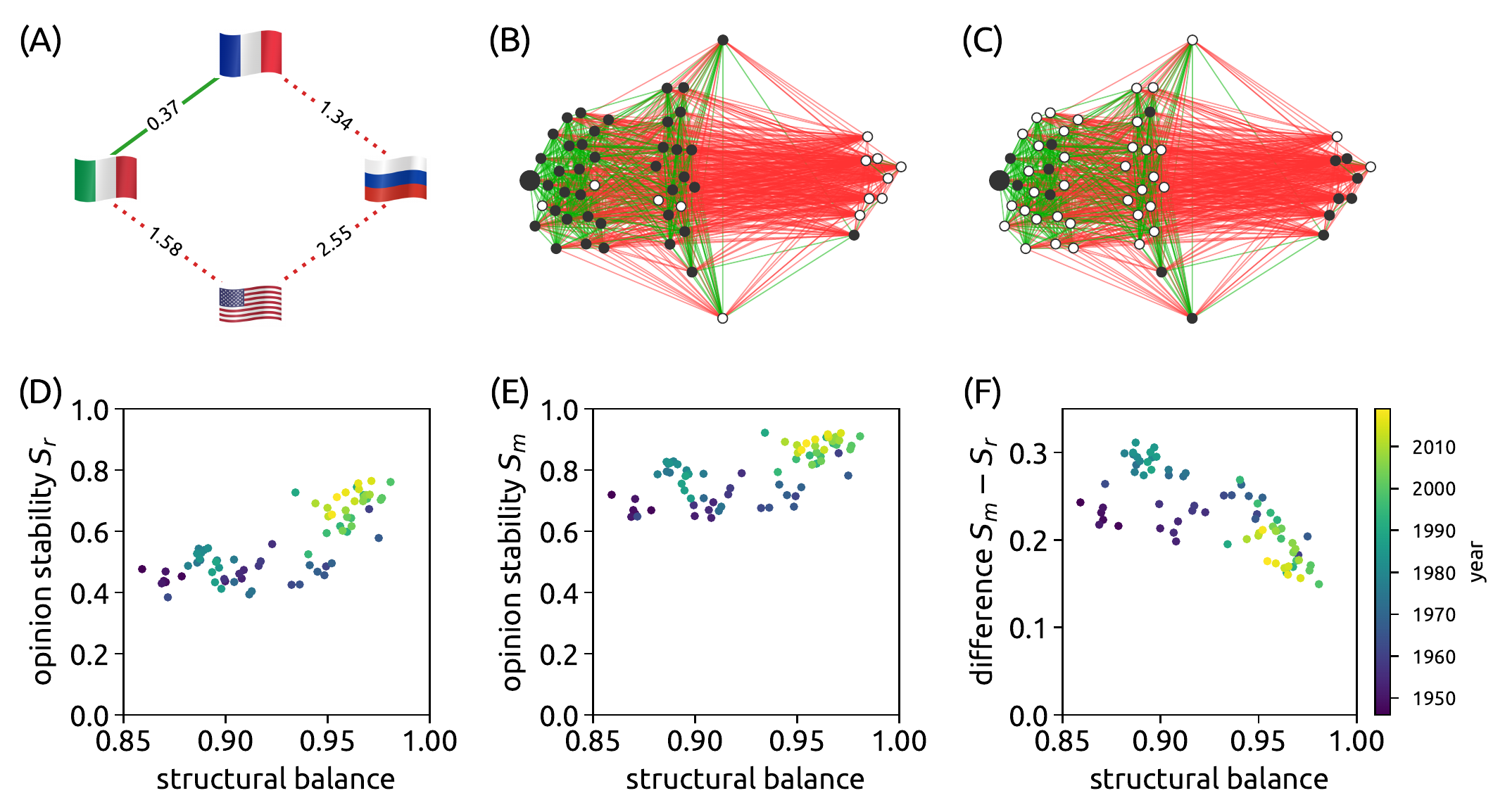}
\caption{Opinion stability for the United Nations General Assembly voting data.
(A) A sample unbalanced subgraph of the last session's dataset.
(B, C) Resulting opinions in two different realizations of the random neighbor rule on the first session's dataset (the seed node is indicated with a larger marker).
(D--F) Opinion stability and the level of structural balance in datasets from individual UN sessions. While panels (B) and (C) show results for the random neighbor rule ($S_r$) and the majority rule ($S_m$), respectively, panel (D) shows the difference.}
\label{fig:UNGA}
\end{figure*}

Equipped with the opinion stability metric, we can assess opinion formation in empirical worlds represented by empirical signed networks. We first use signed networks derived from United Nations General Assembly (UNGA) votes in individual sessions, where countries that vote similarly are connected with positive links and countries that vote differently are connected with negative links (see Methods for the data description). Figure~\ref{fig:UNGA}A shows part of the network corresponding to the latest completed UNGA session 74 (2019--2020). The loop is unbalanced as the product of its link weights is $-1$. As a result, the outcome of opinion formation using the random neighbor rule is not deterministic: Assuming a positive seed opinion on Italy, the formed opinion on Russia is negative if it is made using the path ITA-FRA-RUS or positive if it is made using the path ITA-USA-RUS. This outcome variability then directly translates in results shown in Figure~\ref{fig:UNGA}B,C where two different realizations of the random neighbor rule are shown to demonstrate the high variability of opinions despite a high level of structural balance of the respective UNGA network (in this case, $B=0.86$). This agrees with our results in Figure~\ref{fig:results1}B where opinion consistency decreases quickly with $\beta$ and displays large fluctuations. Finally, Figures~\ref{fig:UNGA}D--F show that the majority rule yields substantially more stable opinions and that the stability difference between the random neighbor rule and the majority rule tends to grow as the level of structural balance decreases.

The number of nodes in the UNGA datasets is limited by the number of countries participating in the assembly's voting (the number of nodes grows from 53 in the 1st assembly to 191 in the 74th). To be able to observe the scaling of opinion stability similar to the scaling of opinion consistency in synthetic data (Fig.~\ref{fig:scaling}), we thus use signed trust networks from two popular online services: Slashdot~\cite{kunegis2009slashdot,leskovec2010signed} and Epinions~\cite{leskovec2010signed} (see Methods for the data description). Note that while Slashdot and Epinions are social networks, our model still differs from classical models of opinion formation on social networks as it concerns opinion-making of an observer, not opinion-making of each individual member of the social network. Nodes in the given social networks represent interconnected subjects on which opinions are made.

\begin{figure*}
\centering
\includegraphics[scale=0.68]{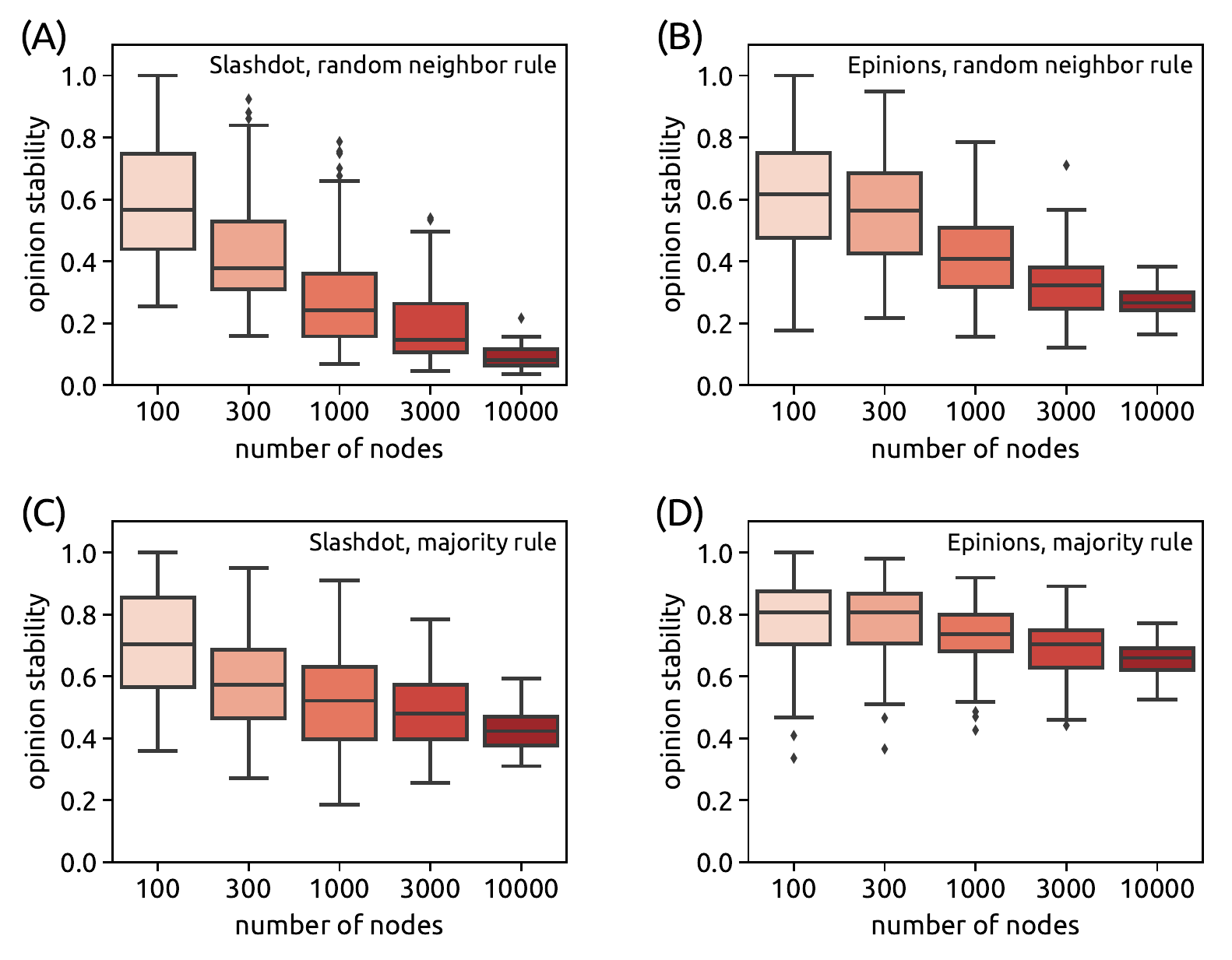}
\caption{Opinion stability achieved in empirical worlds by the random neighbor rule (top) and the majority rule (bottom). Subsets of progressively increasing size have been obtained by sampling from corresponding empirical signed networks. For each subset, we compute stability based on 100 independent seed opinions.\newline
Random neighbor rule: In parallel with the previous results on synthetic relation networks, the opinion stability decreases with $N$ (opinion stability-world complexity tension). The fitted scaling exponents are 0.40 and 0.20 for Slashdot and Epinions, respectively.\newline
Majority rule: While opinion stability vanishes with $N$ slower than it does for the random neighbor rule, the fitted scaling exponents are significantly positive for both Slashdot and Epinions (their respective values are 0.09 and 0.05). Despite the subsets being rather dense (see Fig.~S9 in SM), the opinion stability-world complexity tension is not averted by the majority rule.}
\label{fig:other_real}
\end{figure*}

We create multiple subsets of each network with progressively increasing numbers of nodes (see Methods for details). We find that a stability-complexity tension is present in the real worlds (Fig.~\ref{fig:other_real}, panels D and E): opinion stability consistently decreases with the number of subjects. The fitted scaling exponents are 0.40 and 0.20 for Slashdot and Epinions, respectively. These values cannot be directly compared with the scaling exponent $2\beta$ that we derived for opinion consistency in random networks as Slashdot and Epinions networks are manifestly non-random.
Building on the understanding that we gained by analyzing simulations on synthetic worlds, we can conclude that the levels of noise in the two real relation networks are high, which makes opinion formation using the random neighbor rule unreliable.

We finally apply the majority rule on real networks. In agreement with previous results, we observe that the resulting opinion stability is almost always higher than achieved by the random neighbor rule and the difference generally grows as the network's level of structural balance decreases (see Fig.~S10 in SM). As $N$ increases, the average opinion stability still vanishes with $N$, albeit slower than is the case for the random neighbor rule (see Figure~\ref{fig:other_real}C,D). We can thus again conclude that the majority rule does not solve the fundamental problem identified by our work: the formed opinions become progressively less reliable as the system size grows.

\section*{Discussion}
We demonstrated that sequential opinion formation based on an explicit network of trust and distrust is inherently fragile: even a small amount of noise leads to inherently fragile (\ie, inconsistent and unstable) outcomes. This suggests that to prevent the spreading of misinformation in large-scale online systems, it is paramount that there exist no trust links from credible to low-credibility sources of information. If a tiny fraction of such misleading links exists, an observer who starts trusting credible information sources may end up trusting a substantial number of low-credibility sources. If the same happens for a large number of observers, a cluster of misinformed individuals (such as anti-vaccination clusters~\cite{johnson2020online}) can thrive.

The more complex our world, the more fragile the process -- there is a tension between opinion consistency/stability and the world's complexity. An individual observer can compensate for this increasing fragility by forming an independent (\ie, not derived from the trust network) initial opinion on a larger number of subjects, before resorting to the trust network to form an opinion on the remaining subjects. For example, a person forming opinions about a set of interconnected websites -- some of them based on scientific content, some of them promoting conspiracy material -- can increase the opinion consistency/stability by first carefully evaluating the trustworthiness of a large number of websites, and only then relying on relations between already trusted/distrusted websites to form opinions on the remaining ones. This suggests that policy to increase the consistency/stability of collective opinion may aim to promote the formation of individuals' independent opinion about a substantial number of subjects. Within the studied framework, the majority rule yields better results, yet: (1) the majority rule is more laborious than the random neighbor rule as it assumes collecting all direct evidence on each target node, (2) opinion consistency and stability under the majority rule still vanish when the network is not sufficiently dense and the noise is not sufficiently low.

From the observer standpoint, our work focuses on little cognitively-demanding opinion formation mechanisms, which opens the way to studying more sophisticated mechanisms, and understanding the trade-off between the robustness of the resulting opinions and the observer's cognitive costs required to form the opinions. In the real world, additional influences---social influence, in particular---and heuristics are likely to be at work, and a high level of heterogeneity across observers is expected. Whether additional influences and mechanisms further increase or mitigate the fragility of the individuals' opinion formation process is a priori unclear. More sophisticated opinion formation models and their calibration to empirical data hold promise to shed light on this fundamental process for our interconnected societies.

\clearpage

\section*{Methods}

\subsection{Opinion consistency.}
When the ground truth division of subjects in camps is known, we compute the consistency of a given opinion vector $\vek{o}$ with the ground truth assignment $\vek{T}$ using \eref{consistency}. The seed opinions are excluded from the computation of consistency as these opinions are consistent by construction. The consistency values range from $-1$ for the opinion vector that totally disagrees with the ground truth (except for the seed opinions, the observer has positive opinions on all subjects from camp 2 and negative opinions on all subjects from camp 1) to $+1$ for the opinion vector that perfectly matches the ground truth camps. The consistency of a random opinion vector is zero on average. Note that we assume here that opinions have been eventually made on all subjects, which is the case for all simulations presented here.

In numerical simulations, we average over independent model realizations on multiple realizations of the synthetic two-camp networks to estimate the mean opinion consistency. In Figure~\ref{fig:results1}, we complement the mean with the 10th--90th percentile range of the consistency values. In Figure~\ref{fig:scaling}, we assess the uncertainty of mean consistency with the standard deviation of mean consistency over 10,000 bootstrapped sets of results; the displayed error bars are 3-times of that.

\subsection{Master equation solution.}
\label{sec:ME}
The proposed opinion formation model using the random neighbor rule can be studied analytically for the two-camp relation network. We use the number of formed opinions, $n$, and the number of ``consistent'' opinions, $c$, as the variables to describe the process. Initially, $n=N_S$ and $c=N_S$, because all seed opinions are assumed to be consistent (they are all positive opinions on subjects from camp 1 or, more generally, positive opinions on subjects from camp 1 and negative opinions on subjects from camp 2). We introduce the probability distribution of $c$ consistent opinions after $n$ opinions are formed, $P(c; n)$, with normalization $\sum_c P(c;n) = 1$. The initial condition is $P(N_S;N_S)=1$ in line with the description above. To find $P(c;n)$ for $n>1$, we write the general master equation
\begin{widetext}
\begin{equation}
\label{ME-general}
P(c; n) = P(c - 1; n - 1) W(c-1\to c; n-1) + P(c; n - 1) \big[1 - W(c\to c + 1; n-1)\big].
\end{equation}
\end{widetext}
The transition probability $W(c-1\to c; n-1)$ corresponds to making a consistent opinion in a situation when $n-1$ opinions have been made, of which $c-1$ are correct.

If the target subject (on which the opinion is to be formed) is from camp 1, $W(c-1\to c; n-1)$ is the probability that the observer decides to form a positive opinion on the target subject. Assume now that there are $t_1$ trusted (\ie, with a positive opinion) subjects from camp 1, $d_1$ distrusted (\ie, with a negative opinion) subjects from camp 1, $t_2$ trusted subjects from camp 2, and $d_2$ distrusted subjects from camp 2. The probability of forming a positive opinion on the target node is $n_P/(n_P + n_N)$ where $n_P$ and $n_N$ are the numbers of neighbors of the target subject that---when chosen---would result in forming positive and negative opinions, respectively. The expected value of $n_P$ is proportional to
$$
t_1(1-\beta) + d_1\beta + t_2\beta + d_2(1-\beta) = (c-1)(1-\beta) + (n - c)\beta
$$
where we used that $t_1 + d_2=c-1$ and $t_1+d_1+t_2+d_2=n-1$. Note that it is the random structure of the relation network that allowed us to write a simple expression. Similarly, $n_N$ is proportional to
$$
t_1\beta + d_1(1-\beta) + t_2(1-\beta) + d_2\beta = (c-1)\beta + (n - c)(1-\beta).
$$
Taken together, these formulas give us the transition probability $W(c-1\to c; n-1) = n_P / (n_P + n_N) = [c(1-2\beta)+\beta(n+1)-1] / (n - 1)$. It can be checked easily that the form of $W(c-1\to c; n-1)$ is the same when the target subject is from camp 2. By plugging this $W(c-1\to c; n-1)$ in \eref{ME-general}, we obtain the master equation \eref{ME} which describes how $P(c;n)$ changes as $n$ grows.

Note that the fact that $W(c-1\to c; n-1) = n_P / (n_P + n_N)$ implies that the same master equation---and thus the same opinion formation model---is obtained by a seemingly more thorough observer who first evaluates all neighbors of the target subject and counts the number of subjects whose choice would result in forming a positive and a negative opinion, $n_P$ and $n_N$, respectively. Based on $n_P$ and $n_N$, the opinion on the target subject can be formed in a probabilistic manner: positive with probability $n_P / (n_P + n_N)$ and negative otherwise. The outcome is thus the same as choosing one neighbor of the target opinion at random and forming the opinion accordingly.

\subsection{Opinion stability.}
Opinion consistency assumes that the ground truth division of subjects in camps is known but that is not the case for most real datasets. To overcome this difficulty, we introduce another metric to assess the formed opinions, opinion stability. For a given relation network, we fix the opinions on subject $i$ and use $R$ independent model realizations to compute the average opinion $\overline{o_j}$ for all other subjects. If the formed opinions on subject $j$ are stable, they are the same in all or most realizations and the value $\overline{o_j}$ is thus close to $+1$ or $-1$. By contrast, volatile formed opinions result in $\overline{o_j}$ close to zero. We then compute the average opinion stability with respect to the seed subject $i$ as
\begin{equation}
\label{stability_first}
S_i' = \frac1{N - 1}\sum_{j\neq i} \babs{\overline{o_j}}
\end{equation}
where the absolute value reflects the fact that both $\overline{o_j}=1$ and $\overline{o_j}=-1$ are signs of stable opinions on subject $j$. Note that the seed opinion is again excluded from the summation. Opinion stability is $S'_i=1$ when all realizations yield the same opinion on $i$. For random opinions, however, the stability is not zero due to the absolute value in \eref{stability_first} which is never negative. In that case, $\overline{o_j}$ follows the normal distribution with zero mean and standard deviation $1/\sqrt{R}$. It can be shown that the mean of $\abs{\overline{o_j}}$ is $\sqrt{2/(R\pi)}$ which represents the expected opinion stability of a random trust vector. We thus transform \eref{stability_first} as
\begin{equation}
\label{stability_formula}
S_i = \frac{S_i' - \sqrt{2/(R\pi)}}{1 - \sqrt{2/(R\pi)}}
\end{equation}
to obtain the final formula for opinion stability. Its values range from zero, on average, when opinions on all subjects are random to one when opinions on all subjects are the same in all model realizations. Individual $S_i$ values can be used to characterize the stability of opinions based on a seed opinion on subject $i$ or aggregated to represent the overall opinion stability. In simulations on synthetic relation networks, we use 100 independent network realizations and compute opinion stability for a randomly chosen node. In simulations on real relation networks, we present opinion stability results for 100 nodes chosen at random. See Sec.~S4 in SM for a comparison between opinion consistency and opinion stability.

\subsection{Real datasets}
We test the opinion formation model on three distinct real datasets.

The UNGA dataset contains the votes by countries at United Nations General Assemblies~\cite{bailey2017estimating}.\footnote{The data are available at \url{https://dataverse.harvard.edu/dataset.xhtml?persistentId=doi:10.7910/DVN/LEJUQZ}.}
We use the state ideal point positions in one dimension estimated in~\cite{bailey2017estimating} from the voting data to quantify ``state positions toward the US-led liberal order''. The dataset contains all 74 general assemblies held in the years 1946--2020; assembly 19 is ignored because of faulty data. For each assembly, estimated state positions $x_i$ can be directly translated in distances $\lvert x_i - x_j\rvert$ between the states. We generate one signed network for each general assembly by first removing all countries with less than 20 votes (up to 7 countries have been removed in one session) and then representing state distances below the 33.33th percentile (for the given general assembly) as positive links and state distances above the 66.67th percentile as negative links, saving the network's giant component. The numbers of nodes and links in the network increase progressively from 53 and 919, respectively, in the 1st general assembly to 191 and 12,096, respectively, in the 74th. The numbers of positive and negative links are identical by construction in each network; the level of structural balance ranges from 0.86 to 0.98.

The Slashdot dataset represents the social network website social network where the users can tag each other as friends or foes~\cite{kunegis2009slashdot,leskovec2010signed}.\footnote{The data are available at \url{http://snap.stanford.edu/data/soc-sign-Slashdot090221.html}.}
While the original network is not symmetric, we represent it as symmetric, neglecting the mutual links whose signs do not agree (less than 1\% of all links) and finally keeping only the giant component. The resulting Slashdot network comprises 82,052 nodes and 498,527 signed links.
The fraction of negative links is $0.236$ and the network's level of structural balance is $B=0.867$.

The Epinions dataset represents the social trust network of the website's users~\cite{leskovec2010signed}.\footnote{Obtained from \url{http://snap.stanford.edu/data/soc-sign-epinions.html}.} After the same processing as we apply to the Slashdot data, the resulting Epinions network comprises 119,070 nodes and 701,569 singed links. The fraction of negative links is $0.168$ and the network's level of structural balance is $B=0.905$.

To study the dependence of results on the network size, we created small subsets of the large Slashdot and Epinions networks by choosing a random node and gradually including its nearest neighbors, second-nearest neighbors, and so on, until a target number of nodes is reached. We created 100 independent networks for each network size, each of them starting from a node chosen at random. An alternative construction by choosing a given number of nodes or links at random would produce very sparse networks whose sparsity would directly impact the opinion formation process (see Fig.~S7 in SM).

\begin{acknowledgments}
This work is supported by the National Natural Science Foundation of China (Nos. 11622538, 61673150, 11850410444). MSM acknowledges financial support from the URPP Social Networks at the University of Zurich, the Swiss National Science Foundation (Grant No. 200021-182659), and the UESTC professor research start-up (Grant No. ZYGX2018KYQD215). LL acknowledges the Science Strength Promotion Programme of UESTC (Grant No. Y030190261010020). Simulation code to reproduce the presented results is available at \url{https://github.com/8medom/OpinionFormation}.
\end{acknowledgments}

\bibliography{refs_trust}

\clearpage

\onecolumngrid
\renewcommand\thefigure{S\arabic{figure}}
\setcounter{figure}{0}
\renewcommand{\thesection}{S\arabic{section}}
\setcounter{section}{0}
\renewcommand{\thesubsection}{S\arabic{section}.\arabic{subsection}}
\setcounter{subsection}{0}
\renewcommand{\theequation}{S\arabic{equation}}
\setcounter{equation}{0}

\begin{center}
\Large\bfseries The fragility of opinion formation in a complex world\\
Supporting Material
\end{center}

\section{Opinion formation on a balanced relation network}
\label{sec:special_case}
In the two-camp case, $\beta=0$ leads to the ideal two-camp structure without any noisy links. Assuming that a positive seed opinion on a subject from camp 1, the resulting opinions are positive for all subjects from camp 1 and negative for all subjects from camp 2. The opinion consistency and stability are then one. A more general proposition is as follows: When all loops in the relation network are balanced, the opinions formed from a single seed opinion are deterministic. The notion of loop balance is a direct generalization of balanced and imbalanced triads: A loop is balanced if the product of edge signs along the loop is one. An imbalanced loop has the product of edge signs along the loop equal to $-1$. To prove this proposition, consider loop $L$ that contains the seed node, $s$ and take any other node, $i$, in this loop. The loop can be now split in two independent paths, $\Gamma_1$ and $\Gamma_2$, leading from $s$ to $i$. According to the probabilistic rule, opinion $o_i$ formed using path $\Gamma$ from $s$ to $i$ can be written as $o_i = o_s\prod_{e\in\Gamma} R(e)$ where $R(e)$ is the sign of edge $e$ in the relation network. Since $\Gamma_1\cup\Gamma_2=L$, we can write
$$
\prod_{e\in L} R(e) = \big(\prod_{e\in\Gamma_1} R(e)\big) \times \big(\prod_{e\in\Gamma_2} R(e)\big)
$$
Now if loop $L$ is balanced, then $\prod_{e\in L} R(e)=1$ on the left-hand side. As a result, the two terms on the right-hand side must have the same sign which then directly implies that $o_i$ is the same regardless of whether $\Gamma_1$ or $\Gamma_2$ are used to form the opinion on node $i$---the opinion-formation outcome is deterministic when only loop $L$ is used. If all loops in the relation network are balanced, then the outcome is deterministic regardless of which paths are used to form the opinions. When all opinions formed using a given seed opinion are always the same, the opinion stability metric is one. Conversely, if at least one imbalanced loop is present in the relation network, then there is some randomness in the resulting opinions and the resulting opinion stability is less than one. Finally, note that it is possible that all triads in the network are balanced (hence the level of structural balance, $B$, is one), yet some longer loops are imbalanced (see  G. Facchetti, G. Iacono, C. Altafini, Computing global structural balance in large-scale signed social networks, PNAS 108, 20953, 2011 for how to evaluate global structural balance in a signed network).

\section{Solution of the master equation for synthetic two-camp relation networks}
As described in the main text, the opinion formation model can be analyzed in terms of the number of consistent opinions, $c$, and the number of formed opinions, $n$. For the two-camp relation network, we derived the master equation
\begin{equation}
\label{SI_ME}
P(c; n) = P(c - 1; n - 1)\,\frac{c(1-2\beta) + \beta(n+1)-1}{n - 1} + P(c; n - 1) \bigg[1 - \beta - \frac{c(1-2\beta)}{n-1}\bigg]
\end{equation}
which describes how the probability distribution $P(c;n)$ relates to the ``previous step'' probability distributions $P(c-1;n-1)$ and $P(c;n-1)$. Beyond solving \eref{SI_ME} numerically, we study the properties of its solution analytically. While the first idea is to study the limit distribution of the fraction of correct opinions, $c/N$, such a solution does not exist as the simulations show that the variance of this distribution goes to zero as $N$ grows (see Figures~2C and 3 in the main text). The decrease of the standard deviation of consistency, $\sigma_C(N)$, albeit very slow when $\beta$ is small, means that a limit distribution does not exist as $C$ approaches a fixed value in the thermodynamic limit.

Since the limit behavior of \eref{SI_ME} cannot be studied, we focus instead of computing the first and second moment of $c$. To this end, we first multiply \eref{SI_ME} with $c$ and sum it over $c=1,\dots,N$ to obtain
\begin{equation}
\label{recurr1}
\avg{c(n)} = \frac{n-2\beta}{n-1}\avg{c(n-1)} + \beta
\end{equation}
which relates $\avg{c}$ in two consequent steps of the model. The initial condition of \eref{SI_ME} is $\avg{c(1)} = 1$ (the first seed opinion is by definition correct). When $\beta=0$, this equation is solved by $\avg{c(n)}=n$ as expected. When $\beta=0.5$ (when the two-camp structure ceases to exist), the solution is $\avg{c(n)} = (n + 1) / 2$. \eref{recurr1} can be also solved in general, leading to
\begin{equation}
\avg{c(n)}=\frac{n}{2}+\frac{\Gamma(n+1-2\beta)}{2\Gamma(2-2\beta)\Gamma(n)}.
\end{equation}
The leading order contribution to the second term is $n^{1-2\beta}/\Gamma(2-2\beta)$.

When $c$ of $n$ opinions are correct, the remaining $n-c$ opinions are not correct. Since the seed opinion is not included in the evaluation of opinion consistency defined in Eq.~(2) in the main text, the corresponding consistency can be thus written as $[c - 1 - (n - c)] / (n - 1) = (2c-n-1) / (n-1)$. When $c=n$, we obtain $C=1$. By contrast, when $c=1$ (only the seed opinion is correct), we obtain $C=-1$ (maximally inconsistent opinions). The obtained result for $\avg{c(n)}$ can be thus used to find the average opinion consistency as $\avg{C(n)} = [2\avg{c(n)}-n-1]/(n-1)$ whose leading contribution is in turn $\avg{C(n)} = 1 / [\Gamma(2-2\beta)n^{2\beta}]$ which agrees with the simulation results in Figure~3 in the main text.

The behavior of $\avg{c^2(n)}$ can be studied analogously. When \eref{SI_ME} is multiplied with $c^2$ and summed over $c=1,\dots,N$, we obtain
\begin{equation}
\label{recurr3}
\avg{c^2(n)} = \left(1+\frac{2 - 4\beta}{n - 1}\right)\,\avg{c^2(n - 1)} + \frac{2\beta(n-2)+1}{n-1}\,\avg{c(n-1)} + \beta.
\end{equation}
We simplify the notation by introducing $\avg{c(n)}:=F_n$ and $\avg{c^2(n)}:=S_n$ (here $F$ and $S$ stand for the first and the second moment, respectively). By subtracting the second power of \eref{recurr1} from \eref{recurr3}, we obtain
\begin{equation}
\label{recurr4}
V_n = \left(1+\frac{2 - 4\beta}{n - 1}\right)\,V_{n-1} - \left(\frac{1 - 2\beta}{n-1}\right)^2\,F_{n-1}^2+
\frac{(1- 2\beta)^2}{n-1}\,F_{n-1}+\beta(1-\beta)
\end{equation}
where $V_n:=S_n - F_n^2$. Once we know $V_n$, equation $C=(2c-n-1)/(n-1)$ implies that the standard deviation of consistency can be found as $\sigma_C(n)=2\sqrt{V_n}/(n-1)$.

\begin{figure}[h!]
\centering
\figSI{0.8}{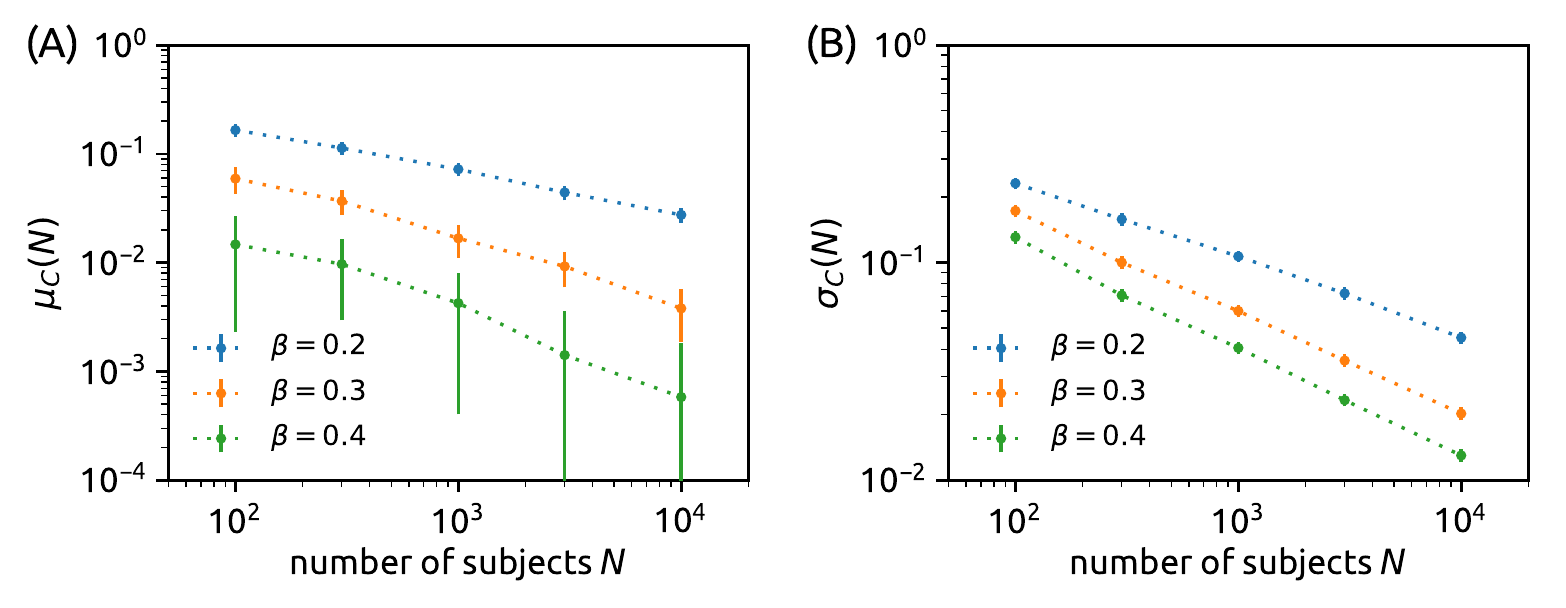}
\caption{Scaling of $\mu_C(N)$ and $\sigma_C(N)$ with $N$ for large $\beta$ (all simulation details as in Figure~3 in the main text). Slopes of fits in the range $1,000$--$10,000$ are 0.42, 0.65, and 0.85, respectively, for mean opinion consistency (panel A) and 0.37, 0.47, and 0.49, respectively, for opinion consistency standard deviation (panel B). The slopes are consistent with the analytical solutions presented in the text.}
\label{fig:higher_beta}
\end{figure}

Using the previously derived solution for $F_n$, \eref{recurr4} can be solved in general, leading to
\begin{equation}
\label{Vn}
V_n = \frac14\left(\frac{\Gamma(n+2-4\beta)}{(1-4\beta)^2\Gamma(1-4\beta)\Gamma(n)}-\frac{\Gamma(n+1-2\beta)^2}{\Gamma(2-2\beta)^2\Gamma(n)^2}-\frac{n}{1-4\beta}\right).
\end{equation}
The leading contributions of the first two terms are $An^{2-4\beta}$. The last term is linear in $n$ which becomes the leading contribution when $\beta>1/4$. When $\beta\leq 1/4$, $\sigma_c\sim n^{1-2\beta}$ and consequently $\sigma_C\sim n^{-2\beta}$. When $\beta>1/4$, $\sigma_c\sim n^{1/2}$ and consequently $\sigma_C\sim n^{-1/2}$. This is confirmed by Figure~\ref{fig:higher_beta} where the scaling of $\mu_C\sim N^{-2\beta}$ for all $\beta$ values but $\sigma_C\sim N^{-1/2}$ for $\beta>1/4$.

\subsection{More seed opinions}
Eqs.~(\ref{recurr1}) and (\ref{recurr4}) hold also in the general case of $N_S$ seed opinions where the initial conditions become $\avg{c(N_S)}=N_S$ (all $N_S$ seed opinions are correct) and $V_{N_S} = 0$ (at $n=N_S$, we surely have $N_S$ correct opinions, hence the variance is zero), respectively. The solution of \eref{recurr1} then has the form
\begin{equation}
\label{Cn_general}
\avg{c(n)} = \frac12\bigg[n+\frac{\Gamma(n+1-2\beta)\Gamma(N_S+1)}{\Gamma(N_S+1-2\beta)\Gamma(n)}\bigg]
\end{equation}
and the resulting leading term of $\mu_C(N)$ is $N^{-2\beta}\Gamma(N_S+1)/\Gamma(N_S+1-2\beta)$. The scaling of $\mu_C(N)$ with $N$ thus does not change when $N_S>1$. The solution of \eref{recurr4}, using \eref{Cn_general} for $F_n$, has the form
\begin{multline}
\label{Vn_general}
V_n = \frac14\left[\frac{\Gamma(1+N_S)}{\Gamma(n)^2}\left(\frac{(1+N_S-4\beta N_S)\Gamma(n)\Gamma(n+2-4\beta)}{(1-4\beta)\Gamma(N_S+2-4\beta)}-\right.\right.\\
\left.\left.-\frac{\Gamma(n+1-2\beta)^2\Gamma(N_S+1)}{\Gamma(N_S+1-2\beta)^2}\right)-\frac{n}{1-4\beta}\right].
\end{multline}
When $N_S=1$, \eref{Vn_general} simplifies to \eref{Vn}. The leading contribution can be found to be again proportional to $N^{-2\beta}$ (when $\beta\leq 1/4$) or to $N^{-1/2}$ (when $\beta>1/4$).

\clearpage
\section{Numerical simulations on synthetic relation networks}

\subsection{Construction of random synthetic networks with fixed degree}
\label{sec:construction}
The basic two-camp setting assumes that subjects $1,\dots,N/2$ form camp 1 and subjects $N/2+1,\dots,N$ form camp 2. Within the camps, the links are positive with probability $1-\beta$ and negative otherwise. Across the camps, the links are positive with probability $\beta$ and negative otherwise. Here $\beta\in[0,0.5]$ plays the role of a noise parameter: As $\beta$ grows, the distinction between the two camps vanishes. The topology of the network is assumed to be random whereas each node has fixed degree $z$. To achieve this, we assign $z$ ``stubs'' to each node and gradually match nodes with free stubs whilst avoiding the nodes linking to themselves and multiple links between a pair of nodes. It is possible that a small number of stubs cannot be matched at the end; those stubs are discarded.

\subsection{Numerical simulations on random synthetic networks with fixed degree}

\begin{figure}[h!]
\centering
\figSI{0.8}{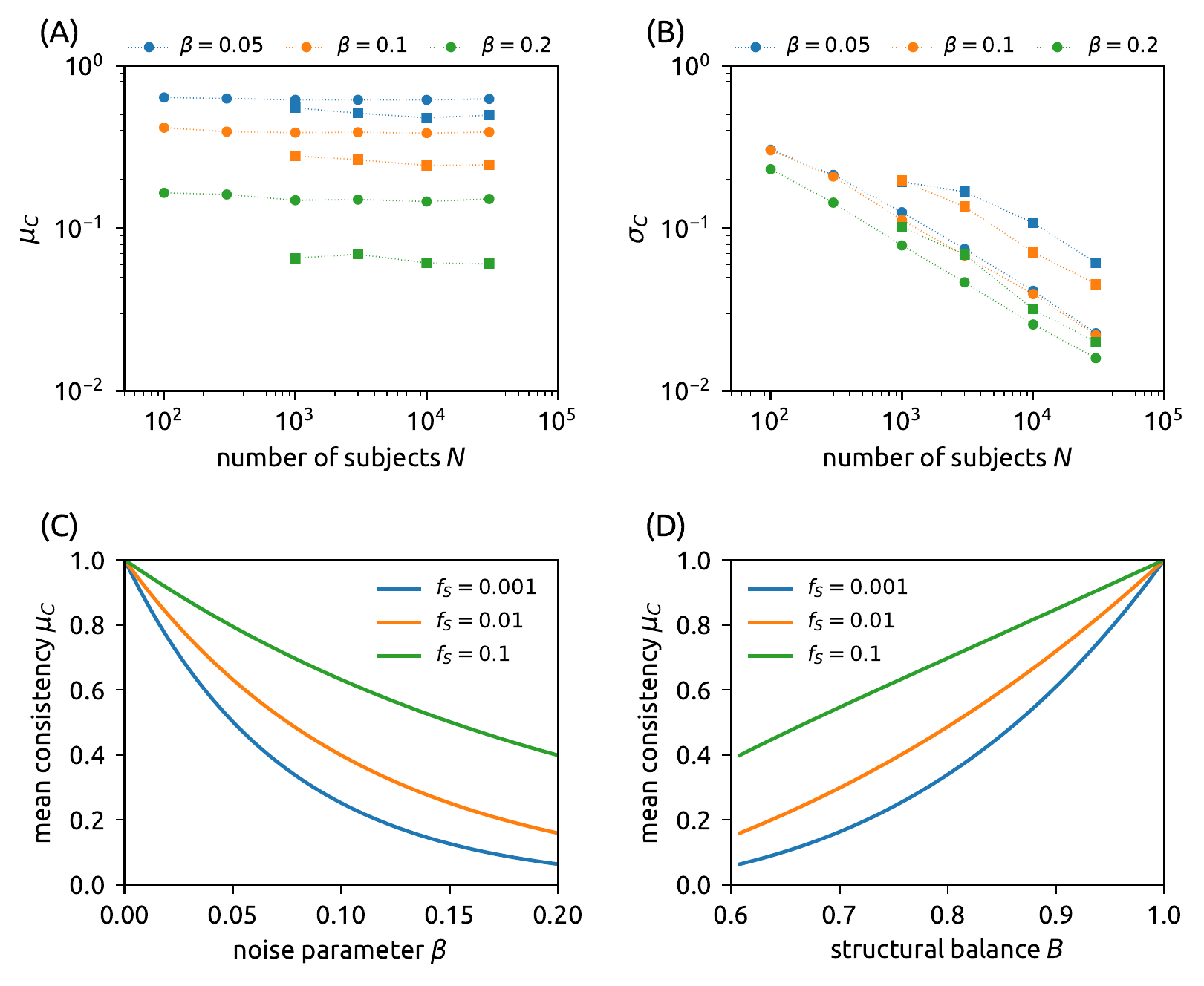}
\caption{Results for a fixed fraction of seed opinions, $f_S$.\newline
(A,B) The scaling of $\mu_C(N)$ and $\sigma_C(N)$ in simulations for $f_S = 0.01$ (circles) and $f_S=0.001$ (squares). We see that regardless of $f_S$ and $\beta$, $\mu_C(N)$ approaches a positive limit value while $\sigma_C(N)$ goes to zero as $N\to\infty$. All parameters as in Figure~3 in the main text. The fitted slope of $\mu_C$ for $f_S=0.01$ and $N$ from $10^3$ are $0$ for all three $\beta$ values. The fitted slope of $\sigma_C$ for $f_S=0.01$ and $N$ from $10^3$ are $-0.50$, $-0.48$, and $-0.47$ for $\beta=0.05$, $\beta=0.1$, and $\beta=0.2$, respectively. These values agree with the derived analytical results.\newline
(C,D) The limit mean consistency $\mu_C=f_S^{2\beta}$ as a function of the noise parameter $\beta$ and the corresponding structural balance $B=(1-\beta)^3 + 3(1-\beta)\beta^2$.}
\label{fig:fixed_seed_fraction}
\end{figure}

Figure~\ref{fig:fixed_seed_fraction} shows that the scaling of opinion consistency with $N$ changes when the number of seed opinions, instead of being fixed, grows with $N$ as $f_SN$. We see that the $\mu_C(N)$ does not converge to zero but to a positive value; the limit of $\sigma_C(N)$ remains the same. To study the limit $N\to\infty$ in this case, we can use the same approach as above. In particular, we can directly use \eref{Cn_general} and plug in $N_S=f_SN$ where $f_S$ is the fraction of seed opinions. The resulting mean consistency $\mu_C(N)$ is given in Eq.~(11) in the main text. The main difference from the previous two cases (one seed opinion and $N_S$ seed opinions) is that the limit $\mu_C(N)$ is not zero in the limit $N\to\infty$: $\mu_C(N)\to f_S^{2\beta}$. For $\sigma_C(N)$ follows
\begin{multline}
(N-1)^2\sigma_C(N)^2=\frac{\Gamma(1+f_SN)}{\Gamma(N)^2}\left(\frac{[1+(1-4\beta)f_SN]\Gamma(N)\Gamma(N+2-4\beta)}{(1-4\beta)\Gamma(f_SN+2-4\beta)}-\right.\\
\left.\frac{\Gamma(N+1-2\beta)^2\Gamma(f_SN+1)}{\Gamma(f_SN+1-2\beta)^2}\right)-\frac{N}{1-4\beta}.
\end{multline}
The leading contribution to $\sigma_C(N)$ can be found to be proportional to $1/\sqrt{N}$. These results are confirmed by the numerical simulations shown in Figure~\ref{fig:fixed_seed_fraction}.

\begin{figure}[h!]
\centering
\figSI{0.8}{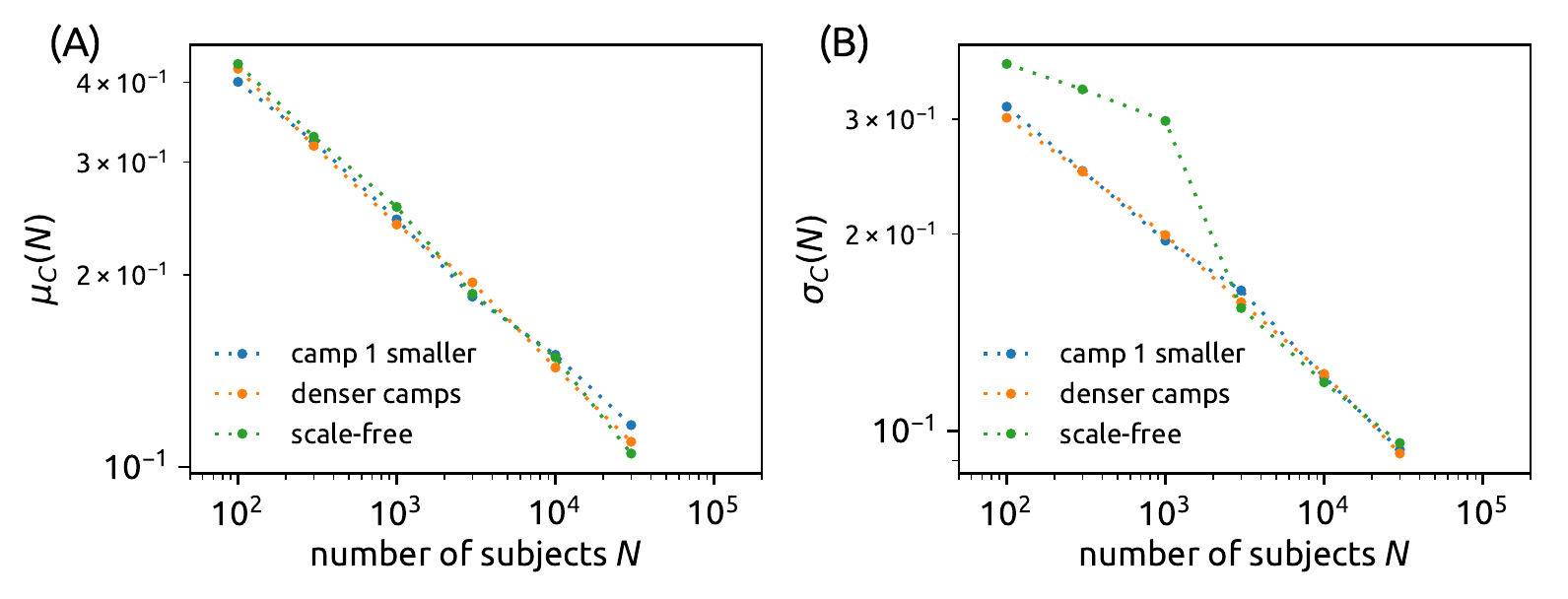}
\caption{The dependence of $\mu_C(N)$ and $\sigma_C(N)$ for two variations of synthetic relation networks with fixed degree. \emph{Camp 1 smaller:} $N/4$ of subjects are in camp 1 and $3N/4$ of subjects are in camp 2. The rest of the construction is the same as in the original two-camp scenario. \emph{Denser camps:} When matching the stubs assigned to the subjects, subject pairs where the subjects belong to different camps are rejected with the probability 70\%, thus leading to links being more frequent within the camps. We see that the results are not significantly influenced by these modifications of the relation networks.}
\label{fig:scaling_variations}
\end{figure}

\subsection{Numerical simulations on synthetic networks with different topologies}
\begin{figure}[h!]
\centering
\figSI{0.8}{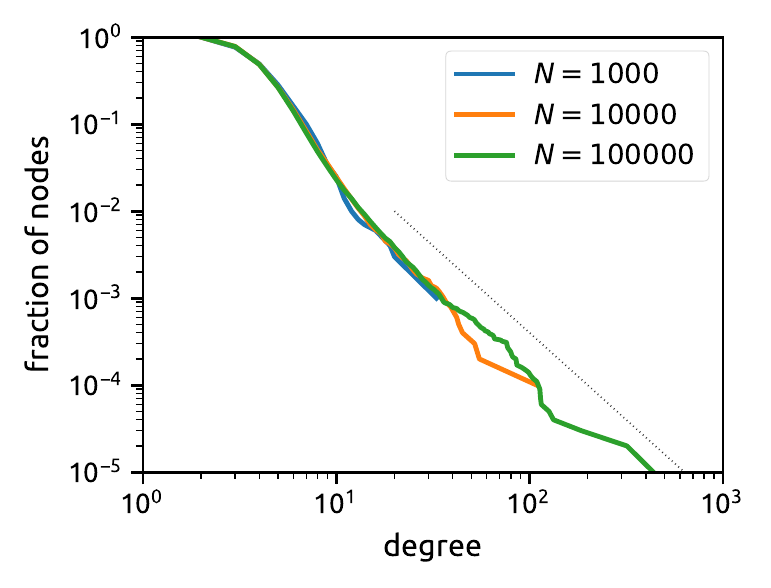}
\caption{Examples of cumulative degree distributions in the described random networks with a power-law degree distribution, $z=4$. The indicative dotted line has the slope of $-2$, corresponding to the the power law exponent $-3$.}
\label{fig:power_law}
\end{figure}

We now finally investigate how the network topology influences the resulting opinion consistency. To this end, we run the model on distinct kinds of synthetic networks:
\begin{enumerate}
\item \emph{Random networks with a power-law degree distribution:} Each node is first assigned two stubs and the remaining $zN - 2N$ stubs are then distributed one by one with probability directly proportional to node ``activity'' value. Node activity values, $a$, are power-law distributed as $1/a^3$ in the range $[1, \infty)$, thus leading to a degree distribution with a power-law tail (see Figure~\ref{fig:power_law}). The rest of the construction is as described in Section~\ref{sec:construction}.

\item \emph{Preferential attachment networks:} Starting with two nodes connected by a link, one new node is introduced in each step and creates $z/2$ links by choosing target nodes with probability directly proportional to their degree. Each existing node can be chosen at most once. In the beginning of the simulation, when the number of available nodes is smaller than $z/2$, the number of created links is adjusted correspondingly. The network is grown until it contains $N$ nodes.

\item \emph{Watts-Strogatz networks:} Starting from a periodic regular 1D lattice where each node has $z$ neighbors, one end of each of the existing links is rewired with the rewiring probability $p_r$. As $p_r$ grows from $0$ to $1$, the resulting networks transition from the regular lattice limit to the random network limit, respectively.
\end{enumerate}
For each kind of networks, their $N$ nodes are assigned at random in two camps of equal size and the link signs are generated in the same way as for the original random networks with fixed degree. All results shown here are for $z=4$ and $\beta=0.1$ (we use here a lower $z$ value to make the heterogeneous degree distribution of preferential attachment networks more pronounced).

\begin{figure}[h!]
\centering
\figSI{0.64}{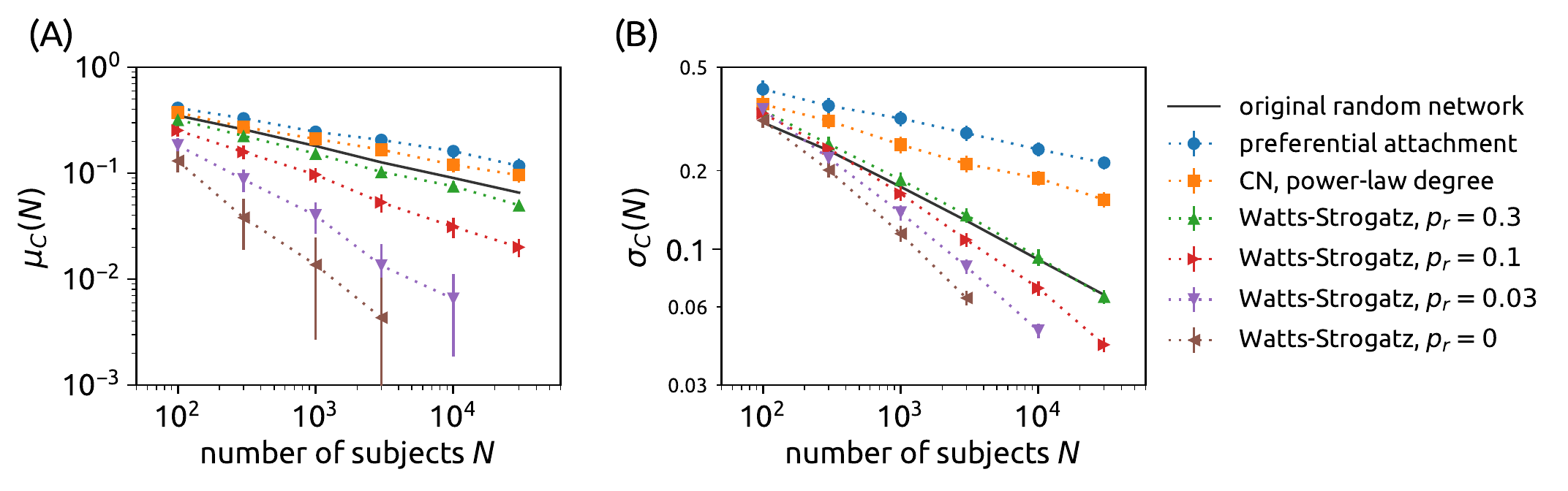}
\caption{The dependencies of $\mu_C$ and $\sigma_C$ on $N$ for $\beta=0.1$. As in Figure~3 in the main text, we run the model once on each of 1,000 network realizations, $z=4$. We see that while the scaling behaviors $\mu_C(N)\sim N^{-\gamma}$ and $\sigma_C(N)\sim N^{-\gamma}$ are maintained for all considered network topologies, the scaling exponent $\gamma$ is strongly influenced by the network topology. In agreement with the comparison presented in Figure~4 in the main text, we see that relationship networks with broad degree distributions (preferential attachment networks and random networks with power-law degree distributions) yield lower exponent $\gamma$ (compared to the original random network with fixed degree) albeit consistency variations are substantially higher and vanish slowly (see panel B). The Watts-Strogatz networks yield significantly higher exponents $\gamma$: the lower the rewiring probability, the faster $\mu_C(N)$ vanishes with $N$.}
\label{fig:topology_scaling}
\end{figure}

\clearpage
\section{Opinion stability vs. opinion consistency}

\subsection{Opinion stability}
\label{sec:stability}
Opinion consistency assumes that the ground truth division of subjects in camps is known but that is not the case for most real datasets. To overcome this difficulty, we introduce another metric to assess the formed opinions, opinion stability. For a given relation network, we fix the opinions on subject $i$ and use $R$ independent model realizations to compute the average opinion $\overline{o_j}$ for all other subjects. If the formed opinions on subject $j$ are stable, they are the same in all or most realizations and the value $\overline{o_j}$ is thus close to $+1$ or $-1$. By contrast, volatile formed opinions result in $\overline{o_j}$ close to zero. We then compute the average opinion stability with respect to the seed subject $i$ as
\begin{equation}
\label{SI_stability_first}
S_i' = \frac1{N - 1}\sum_{j\neq i} \babs{\overline{o_j}}
\end{equation}
where the absolute value reflects the fact that both $\overline{o_j}=1$ and $\overline{o_j}=-1$ are signs of stable opinions on subject $j$. Note that the seed opinion is again excluded from the summation. Opinion stability is $S'_i=1$ when all realizations yield the same opinion on $i$. For random opinions, however, the stability is not zero due to the absolute value in \eref{SI_stability_first} which is never negative. In that case, $\overline{o_j}$ follows the normal distribution with zero mean and standard deviation $1/\sqrt{R}$. It can be shown that the mean of $\abs{\overline{o_j}}$ is $\sqrt{2/(R\pi)}$ which represents the expected opinion stability of a random trust vector. We thus transform \eref{SI_stability_first} as
\begin{equation}
S_i = \frac{S_i' - \sqrt{2/(R\pi)}}{1 - \sqrt{2/(R\pi)}}
\end{equation}
to obtain the final formula for opinion stability. Its values range from zero, on average, when opinions on all subjects are random to one when opinions on all subjects are the same in all model realizations. Individual $S_i$ values can be used to characterize the stability of opinions based on a seed opinion on subject $i$ or aggregated to represent the overall opinion stability. In simulations on synthetic relation networks, we use 100 independent network realizations and compute opinion stability for a randomly chosen node. In simulations on real relation networks, we present opinion stability results for 100 nodes chosen at random.

\subsection{Results of numerical simulations}

\begin{figure}[h!]
\centering
\figSI{0.8}{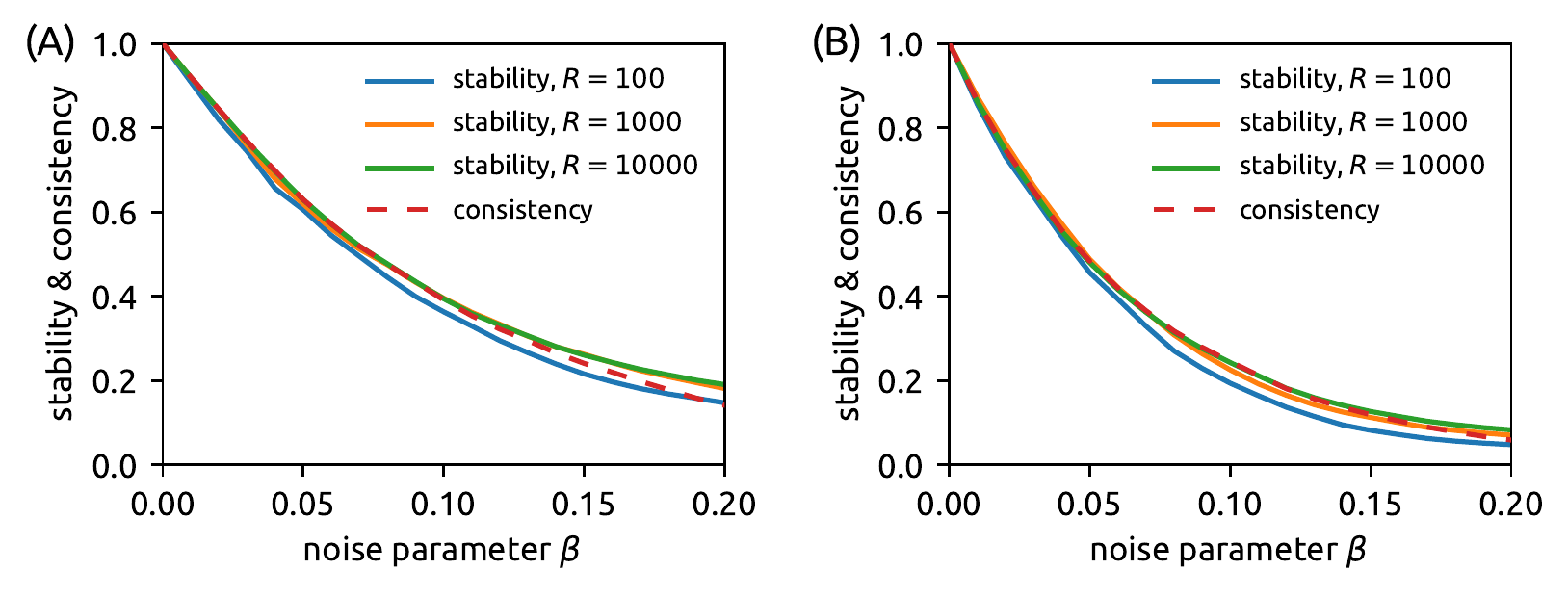}
\caption{Comparing opinion consistency, $C$, and opinion stability, $S$. The two panels show results for $N = 100$ (A) and $N = 1000$ (B). We use one seed opinion and $z=10$ in all simulations. Opinion stability is computed by running $R$ model realizations with a fixed relation network and a fixed seed opinion. The results are then averaged over 100 network realizations. We see that opinion stability weakly depends on the number of realizations, $R$. We use $R=1000$ in all other synthetic data simulations with opinion stability and $R=100$ in the computationally more intensive real data simulations.}
\label{fig:S_vs_beta}
\end{figure}

\begin{figure}[h!]
\centering
\figSI{0.8}{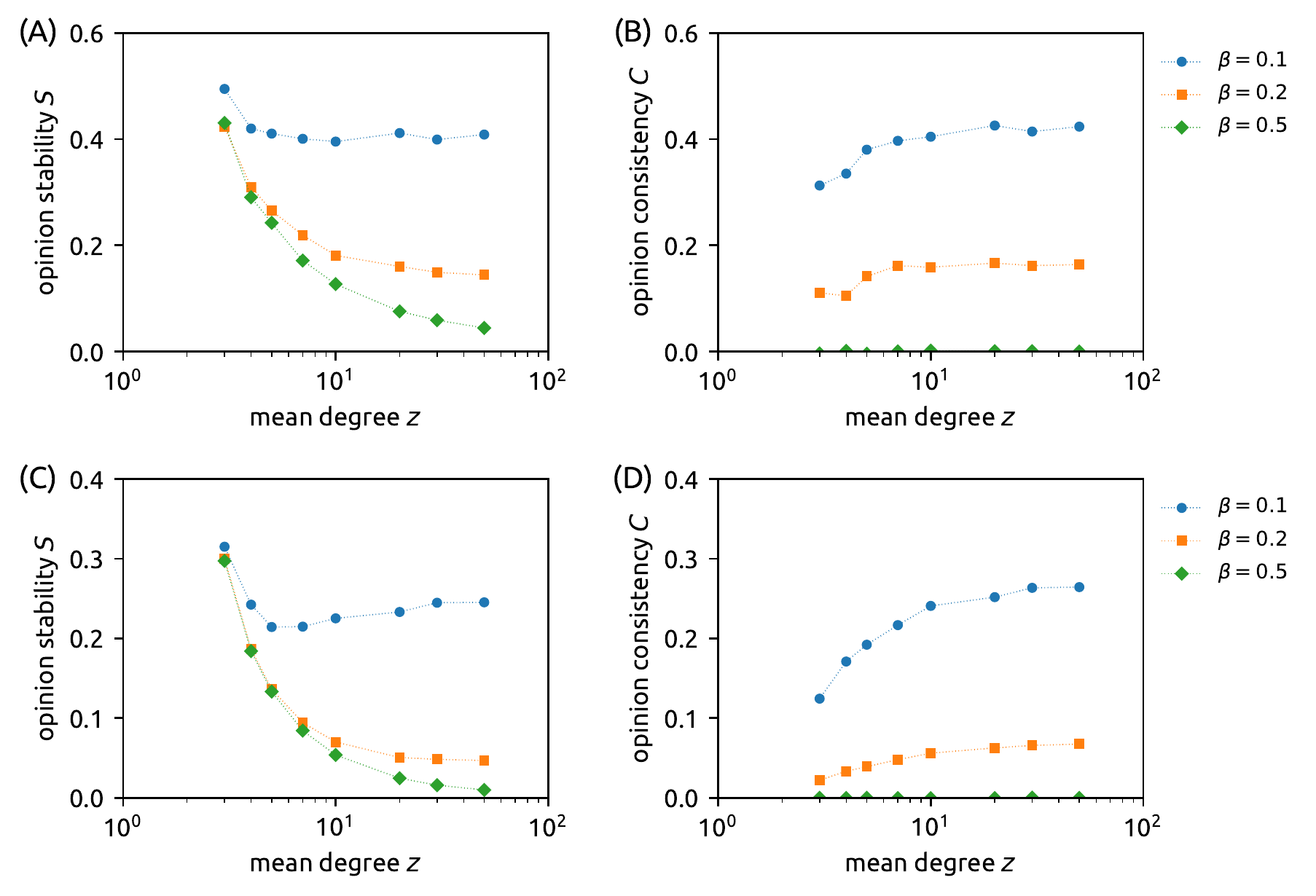}
\caption{Opinion stability and consistency as a function of mean degree $z$ of the two-camp relationship network for $N=100$ (A,B) and $N=1000$ (C,D). Opinion consistency $z$ does, as we have already seen in the main text, depend on $z$, in particular when the noise level is low. The dependence of consistency on $z$ reflects the fact that in a sparse network, a single inverted link in the relation network can lead to the formation of a substantial cluster of opinions inconsistent with the ground truth.\newline
Opinion stability depends on $z$ even stronger and this dependence is counter-intuitive: As $z$ increases (more relation information is added), opinion stability decreases. In fact, $S$ can produce an illusion of stable opinions even when $\beta=0.5$ and the relation network has no structure at all (note that this is properly acknowledged by opinion consistency, $C$, which is then zero for all $z$ values). To understand this apparent opinion stability, consider a tree-like relation network. On such a network, the opinions formed for a given seed opinion are always the same which in turns yields $S=1$. However, these ``stable'' opinions do not reflect any specific structure in the relation network other than the relation network being a tree. Opinion stability therefore becomes a meaningful measure only when the mean network degree is $z\gtrsim 10$. At that point, the relation network makes it possible for the opinion formation process to have various outcomes and a stable opinion on a given subject thus becomes truly informative.}
\label{fig:S_vs_z}
\end{figure}

\begin{figure}[h!]
\centering
\figSI{0.8}{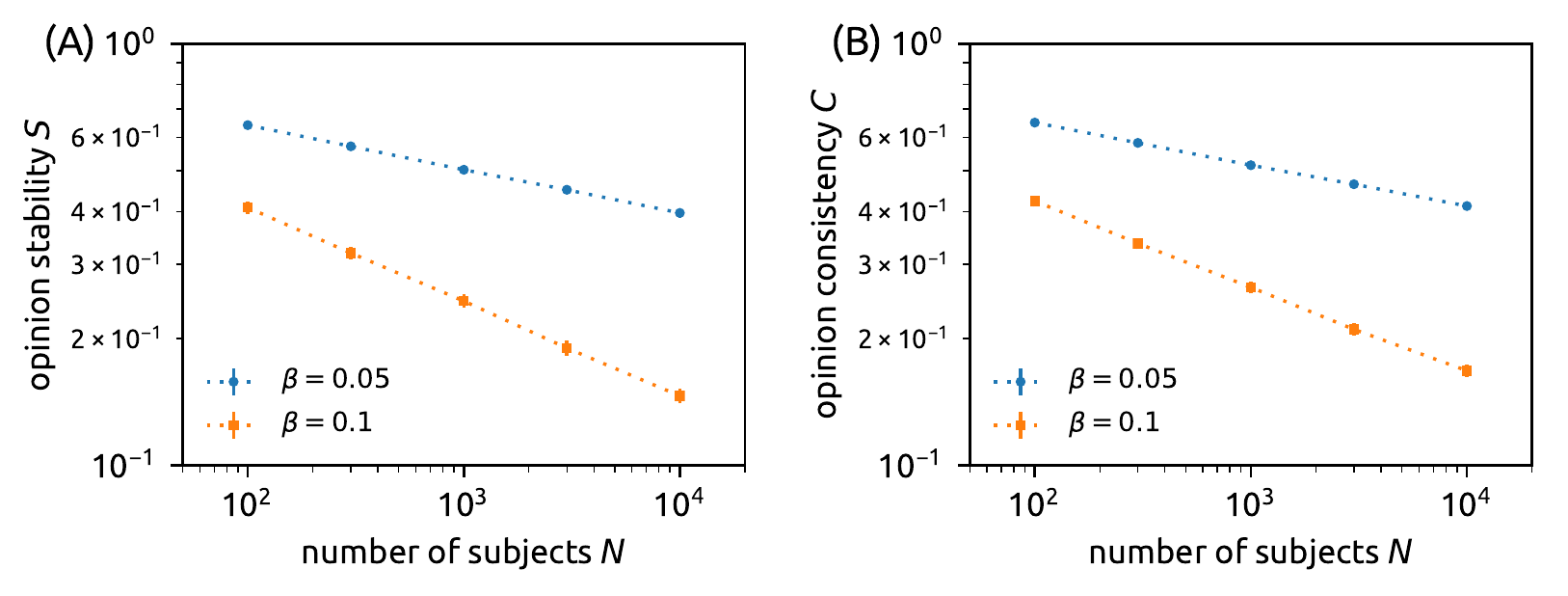}
\caption{As we have already seen for opinion consistency, opinion stability too decreases to zero as $N$ grows. Numerical simulations suggest that both quantities vanish with $N$ with the scaling exponent $2\beta$. The linear fits in the log-log plane in the range 1,000--10,000 are 0.10 for both $S$ and $C$ when $\beta=0.05$. The linear fits in the log-log plane in the range 1,000--10,000 are 0.22 and 0.20, respectively, for $S$ and $C$ when $\beta=0.10$. The results shown here are averaged over 100 independent relation networks with the uniform degree $z=50$ and one seed opinion. Opinion stability is computed from 1,000 independent model realizations for each relation network (the seed opinion is the same in each of these realizations).}
\label{fig:S_vs_N}
\end{figure}

\clearpage
\section{Results on real signed networks}

\subsection{Real datasets}


\begin{figure}[h!]
\centering
\figSI{0.8}{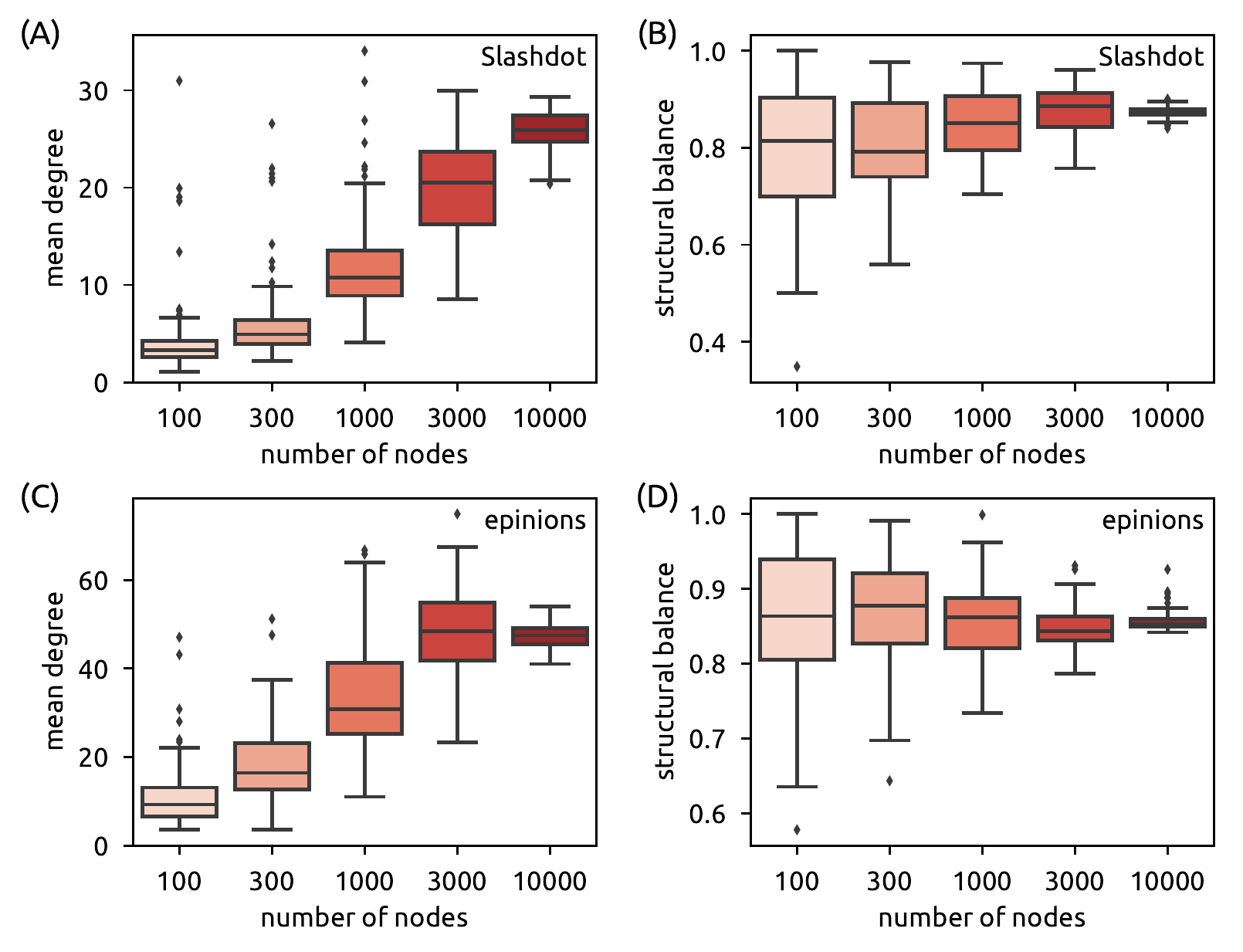}
\caption{The dependencies of the mean degree (A,C) and the structural balance (B,D) on the number of nodes in the created subsets of the Slashdot data (top row) and the Epinions data (bottom row).}
\label{fig:subsets_real}
\end{figure}

\begin{figure}[h!]
\centering
\figSI{0.8}{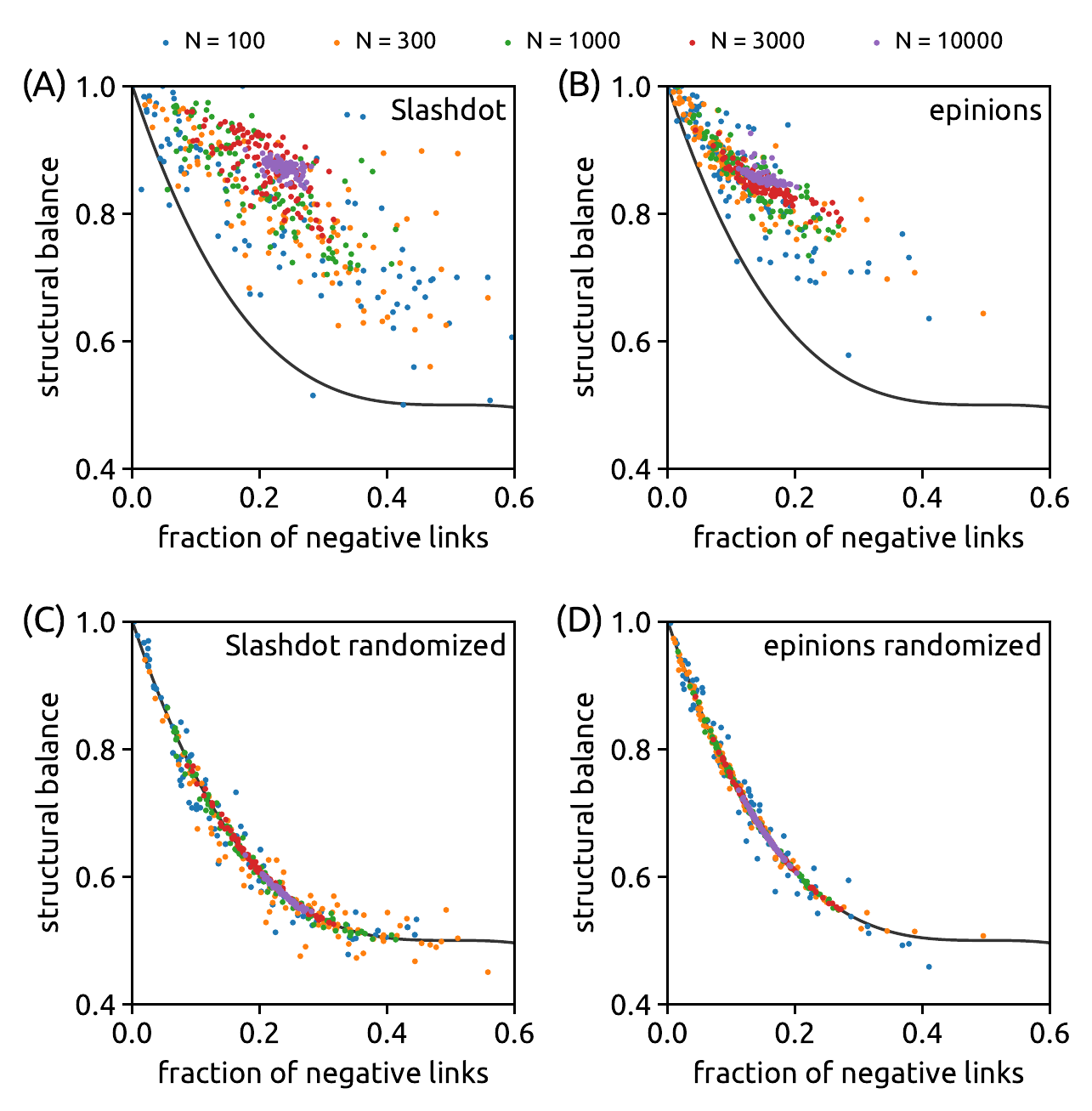}
\caption{The relation between the fraction of negative links and the structural balance, $B$, in the created subsets for Slashdot (A) and Epinions (B); each symbol corresponds to one subset with colors indicating the number of nodes in the subsets. We see here that the vast majority of subsets have structural balance significantly higher than the expected structural balance when negative links are distributed in the networks at random: $B_0 = (1 - f_N)^3 + 3(1-f_N)f_N^2$ (a triad is balanced if all three links are positive or if any of the three links is positive and the other two are negative); this expected value is shown with the solid lines. If the relation signs are randomized (keeping the fraction of negative links fixed), we obtain results in the bottom row where the observed $B$ values are narrowly distributed around $B_0$.\newline
The above observations can be formalized by evaluating the significance of the observed structural balance values by comparing them with the mean and standard deviation of structural balance on networks with randomized signs. For the complete signed networks, the corresponding $z$-scores are $187$ for the Slashdot network (82,052 nodes and 498,527 links, 23.6\% negative links) and 254 for the Epinions network (119,070 nodes, 701,569 links, 16.8\% negative links), respectively.}
\label{fig:triads}
\end{figure}

\subsection{Results of numerical simulations}

\begin{figure}[h!]
\centering
\figSI{0.8}{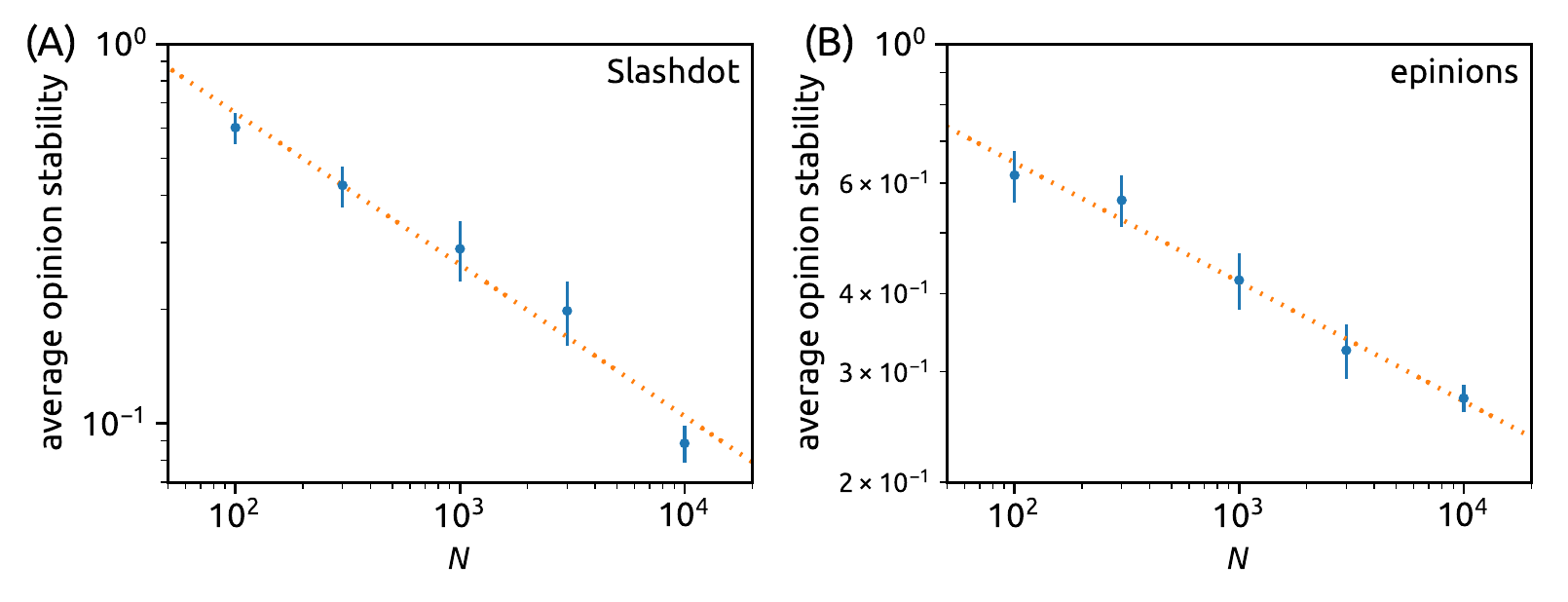}
\caption{The dependence of the mean opinion stability on the subset size $N$ (in a different form, these results are shown in Figure~4 in the main text). The error bars indicate 3-times of each respective standard error of the mean and the dotted line shows the result of the least-squares fit in the log-log plane. The corresponding scaling exponents (slopes of the fitting lines) are 0.40 (Slashdot) and 0.19 (Epinions), respectively, when the fitting lower bound of 100 is used. As shown in Figure~\ref{fig:real_S_vs_B}, the actual stability values are strongly influenced by the structural balance values of the underlying subsets.}
\label{fig:scaling_real}
\end{figure}

\begin{figure}[h!]
\centering
\figSI{0.8}{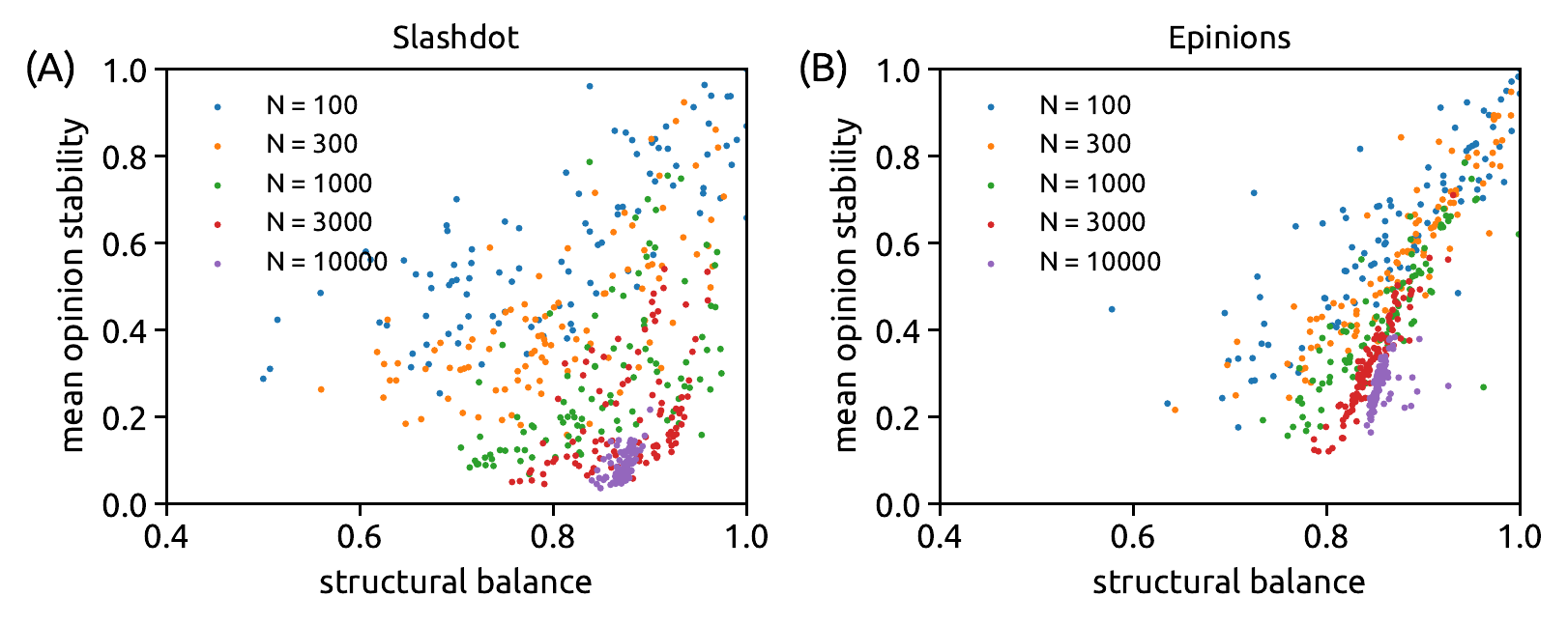}
\caption{The obtained mean stability values for individual subsets in relation to each subset's structural balance (each symbol corresponds to one subset with colors indicating the number of nodes in the subsets). This visualization allows us to disentangle the two major determinants of opinion stability: the subset size, $N$, and the subset structural balance, $B$.}
\label{fig:real_S_vs_B}
\end{figure}

\subsection{The majority rule}

\begin{figure}[h!]
\centering
\figSI{1.}{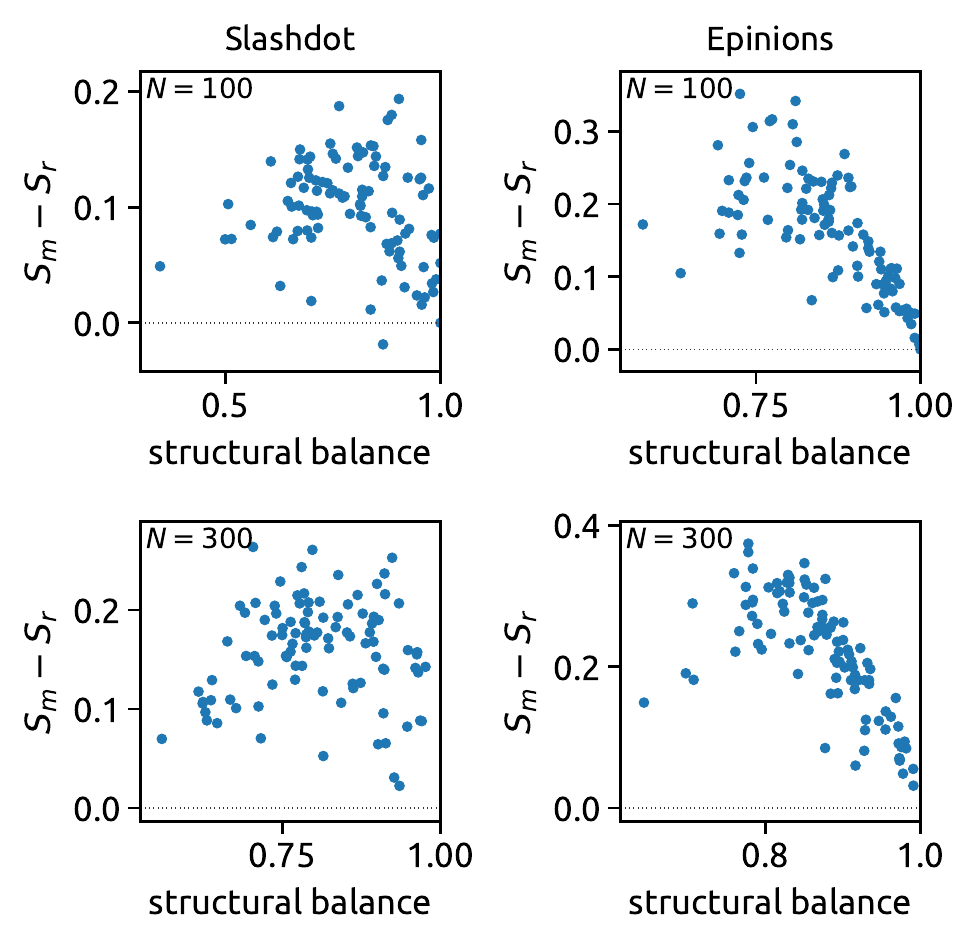}
\caption{The difference between opinion stability achieved by the majority rule, $S_m$, and opinion stability achieved by the probabilistic rule, $S_p$, against the subset's level of structural balance. We show here results for Slashdot (left) and Epinions (right) subsets with 100 nodes (top row) and 300 nodes (bottom row). Of the 400 displayed subsets, only one Slashdot subset with 100 nodes exhibits $S_m - S_p < 0$.}
\label{fig:majo_minus_prob_1}
\end{figure}

\begin{figure}[h!]
\centering
\figSI{1.}{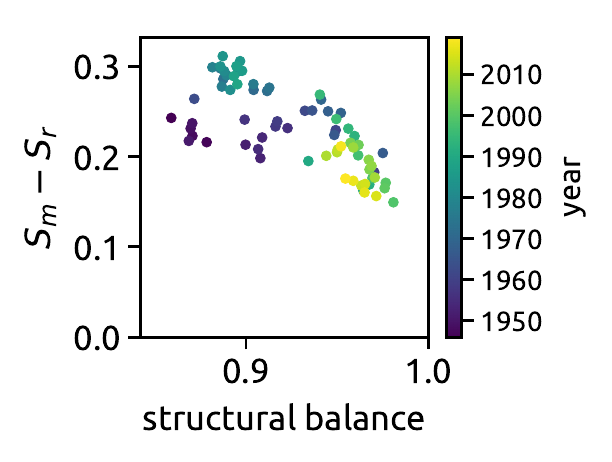}
\caption{As Figure~\ref{fig:majo_minus_prob_1} but for the United Nations General Assembly voting datasets. We use here a color scale to display the year of each general assembly network.}
\label{fig:majo_minus_prob_2}
\end{figure}

\begin{figure}[h!]
\centering
\figSI{0.8}{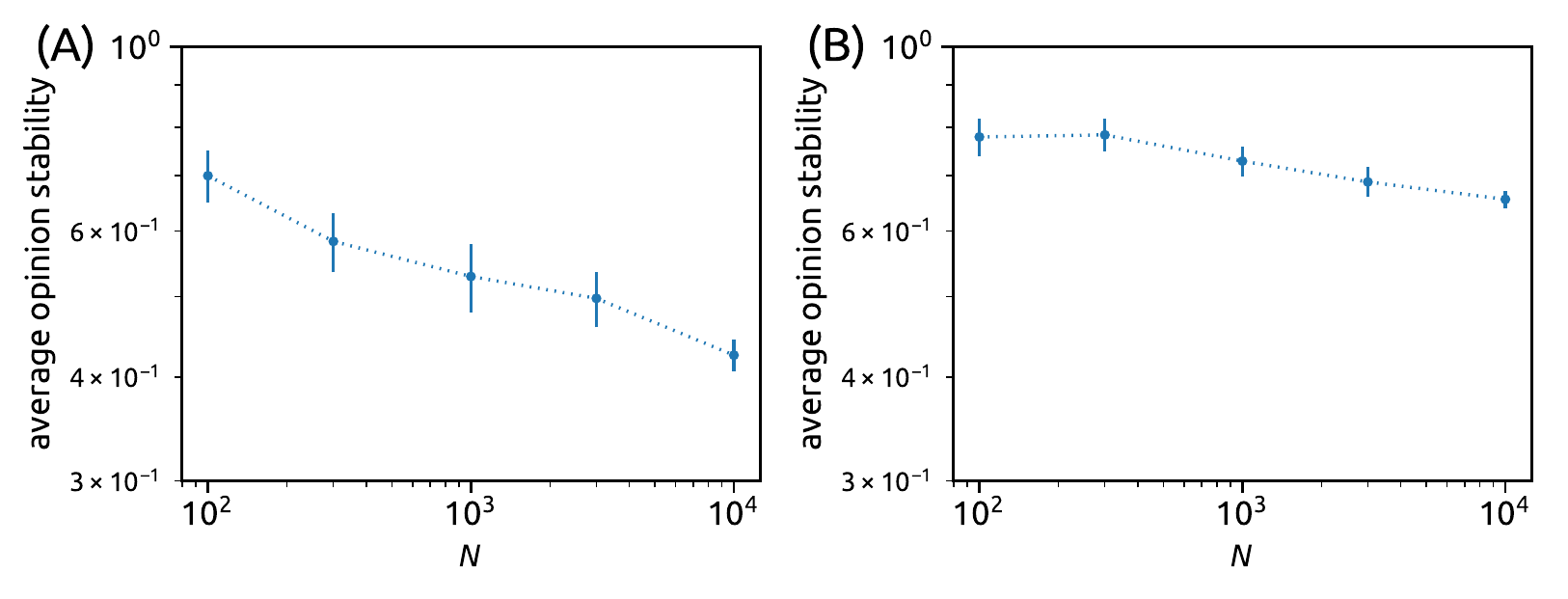}
\caption{The dependence of the mean opinion stability on the subset size $N$ for the majority rule. The error bars indicate 3-times of each respective standard error of the mean and the dotted line shows the result of the least-squares fit in the log-log plane. The corresponding scaling exponents (slopes of the fitting lines) are 0.09 (Slashdot) and 0.05 (Epinions) when the fitting lower bound of 300 is used.}
\label{fig:scaling_real_majority}
\end{figure}

\end{document}